\documentclass[fleqn,usenatbib]{mnras}

\usepackage{newtxtext,newtxmath}

\usepackage[T1]{fontenc}
\usepackage{ae,aecompl}

\usepackage{graphicx}	
\usepackage{amsmath}	
\usepackage{mathtools}
\usepackage{hyperref}
\usepackage{xspace}	
\usepackage{xcolor}	
\usepackage{euscript}
\usepackage{pdflscape}
\usepackage{afterpage}
\usepackage{placeins}
\usepackage{bm}
\usepackage{ dsfont }
\usepackage{listings}
\usepackage{soul}
\usepackage{bbm}

\definecolor{codegreen}{rgb}{0,0.6,0}
\definecolor{codegray}{rgb}{0.5,0.5,0.5}
\definecolor{codepurple}{rgb}{0.58,0,0.82}
\definecolor{backcolour}{rgb}{0.95,0.95,0.92}

\lstdefinestyle{python}{
    backgroundcolor=\color{backcolour},   
    commentstyle=\color{codegreen},
    keywordstyle=\color{magenta},
    numberstyle=\tiny\color{codegray},
    stringstyle=\color{codepurple},
    basicstyle=\ttfamily\footnotesize,
    breakatwhitespace=false,         
    breaklines=true,                 
    captionpos=b,                    
    keepspaces=true,                 
    numbers=left,                    
    numbersep=5pt,                  
    showspaces=false,                
    showstringspaces=false,
    showtabs=false,                  
    tabsize=2,
    upquote=true
}

\newcommand{\gaia}{{\it Gaia}\xspace}
\defcitealias{PaperI}{Paper~I}
\defcitealias{PaperII}{Paper~II}
\defcitealias{PaperIII}{Paper~III}
\defcitealias{PaperIV}{Paper~IV}

\title[Selection function toolbox]{A selection function toolbox for subsets of astronomical catalogues}

\author[D. Boubert and A. Everall]{ 
	Douglas Boubert$^{1,2}$\thanks{E-mail: douglas.boubert@magd.ox.ac.uk}
	and Andrew Everall$^{3}$
	\\
	$^{1}$Magdalen College, University of Oxford, High Street, Oxford OX1 4AU, UK\\
	$^{2}$Rudolf Peierls Centre for Theoretical Physics, Clarendon Laboratory, Parks Road, Oxford OX1 3PU, UK\\
	$^{3}$Institute of Astronomy, University of Cambridge, Madingley Road, Cambridge CB3 0HA, UK\\
}

\date{Accepted XXX. Received YYY; in original form ZZZ}

\pubyear{}

\begin{document}
\label{firstpage}
\pagerange{\pageref{firstpage}--\pageref{lastpage}}
\maketitle

\begin{abstract}
	Large catalogues are ubiquitous throughout astronomy, but most scientific analyses are carried out on smaller samples selected from these catalogues by chosen cuts on catalogued quantities. The selection function of that scientific sample - the probability that a star in the catalogue will satisfy these cuts and so make it into the sample - is thus unique to each scientific analysis. We have created a general framework that can flexibly estimate the selection function of a sample drawn from a catalogue in terms of position, magnitude and colour. Our method is unique in using the binomial likelihood and accounting for correlations in the selection function across position, magnitude and colour using Gaussian processes and spherical harmonics. We have created a new open-source Python package \textsc{selectionfunctiontoolbox} that implements this framework and used it to make three different estimates of the APOGEE DR16 red giant sample selection function, as a subset of 2MASS, with each estimate using an increasing amount of technical knowledge of the APOGEE targeting. In a companion paper we applied our methodology to derive estimates of the astrometric and spectroscopic selection functions of \gaia EDR3. Our framework will make it trivial for astrophysicists to estimate the selection function that they should be using with the custom sample of stars that they have chosen to answer their scientific question.
\end{abstract}

\begin{keywords}
	stars: statistics, Galaxy: kinematics and dynamics, Galaxy: stellar content, methods: data analysis, methods: statistical
\end{keywords}



\section{Introduction}

Much of astrophysics is predicated on testing the predictions of physical theories against a sample of objects drawn from an astronomical catalogue. However, we can only trust these tests if we understand the completeness of our sample. We cannot use a sample to test for the existence of blue horizontal branch stars if blue horizontal branch stars in the catalogue would not have made it into our sample. Neither can we measure the fall off in stellar density away from the Galactic plane, unless we know what portion of the apparent decline is attributable to the incompleteness of our catalogues at faint magnitudes.

The completeness of a sample can be captured in the \textit{selection function}, which for a star (real or counterfactual) with given observed properties $\mathbf{y}$ (often position on the sky, magnitude, and colour) returns the probability that a star in the catalogue would have been included in the sample. The selection function can be strictly one or zero (deterministic cuts that stars either pass or fail) or it can take any value in the unit interval (non-deterministic cuts that stars have some probability of passing). Deducing the true selection function of a sample is impractical because it requires precise knowledge of every detail of the catalogue construction, such as the decisions made in target selection, the individual fields observed, and even knowledge of the seeing throughout the night. Instead, the selection function can be approximated by modelling the catalogue construction \citep{PaperI,PaperII,PaperIII,PaperIV} or by comparing to a full catalogue where one is available. A recent exposition on the difficulties of approximating selection functions was given by \citet{Rix2021}.

Comparison with a full catalogue has previously almost always been performed by taking the ratio of source counts between the sample and catalogue in spatial, apparent magnitude and colour bins \citep{Das2016a,Das2016b,Nandakumar2017,Wojno2017,Chen2018}. The only significant differences between previous approaches is the choice of binning scheme applied. Choosing the right sizes of the bins when using a ratio of counts is important. If the bins are too large then the inferred selection function will be too coarse to capture changes in the true selection function on small scales. If the bins are made too small then bins may contain too few stars for the ratio of counts to be an accurate estimate of the selection function. For spatial bins, a popular choice is to use the field pointings employed by the survey or some form of adaptive mesh refinement \citep{Rybizki2021}. Colour-apparent magnitude bins may be on a regular grid, have data driven boundaries \citep{Mints2019, Bovy2012Segue} or be set by the nominal survey observing strategy \citep{Bovy2014,Stonkute2016,Mackereth2020}. \citet{Everall2020} avoid the over-simplistic count ratio estimation instead using Poisson statistics to forward model the selection function with a smooth Gaussian Mixture Model in colour-apparent magnitude space which avoids binning the data.

Selection functions can usually be split into a known component which is well defined by a survey's documentation (for instance, whether the star lies within one of the fields of a spectroscopic survey) and an unknown component that cannot practically be quantified from the survey's documentation (for instance, the probability that the weather prevented observations of a particular field). In this work we present a general solution to the problem of estimating the unknown component of the selection function of a sample drawn from a catalogue in terms of position, magnitude and colour. Unlike all previous approaches, we use the information that the sample of interest is a \textit{subset} of the full catalogue by treating estimation of the selection function as a binomial statistics problem. We also construct our selection function model to have correlations across sky position, apparent magnitude and colour, solving the problem of small number statistics that has prevented previous works from recovering high resolution selection functions by forcing them to use large bins. We have implemented our methodology in an open-source Python package.

We demonstrate the power of our new methodology by using it to estimate the selection function of the APOGEE DR16 red giant sample as a subset of 2MASS in terms of position and magnitude. In a companion paper we apply this methodology to derive selection functions for the subsets of \gaia EDR3 that have parallaxes, proper motions, and radial velocities.

In Sec. \ref{sec:terminology} we introduce selection functions in the context of a sample drawn from a larger catalogue by the application of cuts. In Sec. \ref{sec:logodds} we explain that the number of stars in the sample in bins across position, magnitude and colour will be binomially-distributed with a number of trials given by the number of catalogued stars in that bin and a probability that is equal to the mean selection probability in that bin. In Sec. \ref{sec:model} we give a detailed account of our model for the selection probability that is correlated across position, magnitude and colour. In Sec. \ref{sec:toolbox} we present an open-source Python implementation of our framework and demonstrate its capabilities and efficacy by estimating the selection function of the APOGEE DR16 red giant sample as a subset of 2MASS.

\section{Methodology}

In this section we lay out our flexible methodology for estimating the selection function of a sample drawn from a catalogue, i.e. what is the probability that a star that entered the catalogue would also enter the sample? A \textit{catalogue} is a list of objects and their properties $\mathbf{y}$, while a \textit{science sample} is a subset drawn from that catalogue. A reader new to this topic should skip to the application of our method to the APOGEE red giant selection function in Sec. \ref{sec:toolbox}, before returning to our highly technical methodology.

\subsection{Selection function of a sample drawn from a catalogue}
\label{sec:terminology}

Suppose we have a science sample of objects drawn from a known catalogue, which were selected according to some cuts $\operatorname{F}(\mathbf{y}) < C$ on functions $\operatorname{F}$ of the catalogued quantities $\mathbf{y}$ of each object where $C$ is a constant which specifies the location of the cut. Examples of such cuts in the form of pairs $(\operatorname{F},C)$ include:
\begin{itemize}
    \item {\it Photometric cuts:} We can select stars brighter than eighteenth magnitude with $(G,18)$, or blue stars with $(G_{\mathrm{BP}}-G_{\mathrm{RP}},0)$.
    \item {\it Spatial cuts:} We can select stars within $10^{\circ}$ of the LMC with $(d_{\mathrm{LMC}}(\alpha,\delta),10^{\circ})$, where $d_{\mathrm{LMC}}(\alpha,\delta)$ computes the great circle distance of a star from the LMC.
    \item {\it Measurement cuts:} We can select stars with a parallax error less than $5\;\mathrm{mas}$ with $(\sigma_{\varpi},5\;\mathrm{mas})$, or a proper motion signal-to-noise ratio greater than five with $(-\mu/\sigma_{\mu},-5)$.
    \item {\it Thinning cuts:} We can select a random 10\% of the catalogue with $(\operatorname{R},0.1)$, where $\operatorname{R}(\mathbf{y})$ is a pseudo-random number generator that uses the catalogued properties as a seed to generate a number between 0 and 1.
\end{itemize}
All ways of selecting subsets of a known catalogue can be written in this form, although the function $\operatorname{F}(\textbf{y})$ may in practise be complicated.

The selection function $\operatorname{S}(\mathbf{y})$ returns the probability that a star with catalogued quantities $\mathbf{y}$ would have satisfied the set of selection cut pairs $\mathcal{S} =\{(\operatorname{F},C)\}$ or, in mathematical form,
\begin{equation}
    \operatorname{S}(\mathbf{y}) = \prod_{\mathclap{(\operatorname{F},C)\in\mathcal{S}}}\operatorname{H}\left(C -  \operatorname{F}(\mathbf{y})\right), \label{eq:known}
\end{equation}
where $\operatorname{H}(x)$ is the Heaviside step function which returns one for positive arguments and zero for negative arguments. Each element of $\mathcal{S}$ specifies a cut used to generate the sample. $\operatorname{S}(\mathbf{y})$ is by definition either zero or unity, implying that if we know the catalogued quantities of a star and the cuts $\mathcal{S}$ then we can say with certainty whether that star is in the sample.

We note the difference between the catalogued quantities $\mathbf{y}$ and the true physical quantities $\mathbf{x}$ of an object. Due to the inevitability of measurement error, the catalogued quantities are always probabilistic transformations of the physical quantities, a relationship which can be captured through a function $\operatorname{P}(\mathbf{y}|\mathbf{x})$ that gives the probability of $\mathbf{y}$ given $\mathbf{x}$. For instance, while the right ascension might appear in both $\mathbf{x}$ and $\mathbf{y}$, the latter is always a noisy measurement of the former. If we wish to know the selection function in terms of the physical quantities, then we must perform the integral,
\begin{equation}
    \operatorname{S}(\mathbf{x}) = \int \operatorname{S}(\mathbf{y})\operatorname{P}(\mathbf{y}|\mathbf{x}) \mathrm{d}\mathbf{y}.
\end{equation}

\subsection{The known and unknown parts of the selection function}
We will generally not want a complicated selection function that depends on all of the catalogued quantities, because if we wish to predict whether a counterfactual star would have made it into our sample then we need to be able to predict each of the catalogued quantities. For instance, if one of the selection cuts requires that a star has a parallax signal-to-noise ratio greater than five, then we need to be able to predict the parallax measurement uncertainty for a counterfactual source. We instead choose a subset $\hat{\mathbf{y}}$ of the catalogued quantities from which we derive an approximate selection function $\operatorname{S}(\mathbf{y})\approx\operatorname{S}(\hat{\mathbf{y}})$.

Suppose we could predict the other catalogued quantities $\tilde{\mathbf{y}}$ through a function $\operatorname{P}(\tilde{\mathbf{y}}|\hat{\mathbf{y}})$ that gives the probability of $\tilde{\mathbf{y}}$ given $\hat{\mathbf{y}}$. Suppose also that we partition the selection cuts into two sets: $\mathcal{S}_{\hat{\mathbf{y}}}$ (those that only depend on $\hat{\mathbf{y}}$) and $\mathcal{S}_{\mathbf{y}}$ (those that depend on at least one quantity in $\tilde{\mathbf{y}}$ but may also depend on $\hat{\mathbf{y}}$). The selection function can then be approximated by marginalising over the quantities $\tilde{\mathbf{y}}$,
\begin{equation}
    \operatorname{S}(\hat{\mathbf{y}}) =  \underbrace{\prod_{\mathclap{(\operatorname{F},C)\in\mathcal{S}_{\hat{\mathbf{y}}}}}\operatorname{H}(C -  \operatorname{F}(\hat{\mathbf{y}}))}_{\operatorname{S}_{\mathrm{known}}(\hat{\mathbf{y}})} \underbrace{\int \mathrm{d}\tilde{\mathbf{y}} \operatorname{P}(\tilde{\mathbf{y}}|\hat{\mathbf{y}}) \prod_{\mathclap{(\operatorname{G},D)\in\mathcal{S}_{\mathbf{y}}}}\operatorname{H}(D -  \operatorname{G}(\mathbf{y}))}_{\operatorname{S}_{\mathrm{unknown}}(\hat{\mathbf{y}})}. \label{eq:knownunknown}
\end{equation}
The elements $(\operatorname{G},D)\in\mathcal{S}_{\mathbf{y}}$ refer to cuts $\operatorname{G}(\mathbf{y})<D$ that depend on at least one quantity in $\tilde{\mathbf{y}}$.

Eq. \ref{eq:knownunknown} can be split into two parts. $\operatorname{S}_{\mathrm{known}}$ is composed of cuts that solely depend on $\hat{\mathbf{y}}$ and which we have perfect knowledge of. Given a star in the catalogue we can say with perfect accuracy whether that star satisfies these cuts. We call the subset of stars in the catalogue that satisfy these cuts the \textit{intermediate sample}. $\operatorname{S}_{\mathrm{unknown}}$ is the expected selection due to the remaining cuts after marginalising over quantities not in $\hat{\mathbf{y}}$. It is likely that we do not understand the catalogue construction sufficiently well enough to know either $\operatorname{P}(\tilde{\mathbf{y}}|\hat{\mathbf{y}})$ nor all of the cuts that were applied, and so must instead find an approximation to $\operatorname{S}_{\mathrm{unknown}}$. \textbf{The objective of this paper is to provide a general methodology for approximating $\operatorname{S}_{\mathrm{unknown}}$ given the stars in the intermediate and science samples.}

We choose to model $\operatorname{S}_{\mathrm{unknown}}(\hat{\mathbf{y}})$ solely in terms of the position, apparent magnitude and \st{apparent} colour of the object. We simplify the modelling of the unknown selection function by assuming that it is piecewise-constant in each of these quantities, i.e. that the unknown selection function is constant within bins in position, magnitude and colour. We assume that there are functions $\operatorname{p}(\hat{\mathbf{y}})$, $\operatorname{m}(\hat{\mathbf{y}})$ and $\operatorname{c}(\hat{\mathbf{y}})$ that map the physical quantities of a star to position $p$, magnitude $m$ and colour $c$ bin indices. To re-iterate, the final selection function will be the product of two functions:
\begin{enumerate}
    \item $\operatorname{S}_{\mathrm{known}}(\hat{\mathbf{y}})$ that incorporates any cuts on quantities $\hat{\mathbf{y}}$ that can be predicted for counterfactual sources and is strictly zero or one,
    \item $\operatorname{S}_{\mathrm{unknown}}(\operatorname{p}(\hat{\mathbf{y}}),\operatorname{m}(\hat{\mathbf{y}}),\operatorname{c}(\hat{\mathbf{y}}))$ that solely depends on quantities $\hat{\mathbf{y}}$ through the position, magnitude and colour bin indices.
\end{enumerate}

\subsection{Binomial likelihood using selection probability}
\label{sec:logodds}

For each star in the science and intermediate samples we compute the tuple of bin indices $(p,m,c)=(\operatorname{p}(\hat{\mathbf{y}}),\operatorname{m}(\hat{\mathbf{y}}),\operatorname{c}(\hat{\mathbf{y}}))$, where $\hat{\mathbf{y}}$ are the catalogued properties of the star that allow us to compute the bin indices. Stars with the same tuple are indistinguishable for our method and so we can summarise their distribution through the quantities $k_{pmc}$ and $n_{pmc}$, which are the number of stars in each $(p,m,c)$ bin in the science and intermediate samples.

If we know the probability $q$ of a star in the intermediate sample in one particular bin\footnote{We have suppressed the indices in this paragraph to aid readability.}, defined by (p,m,c), making it into the sample, then the binomial probability mass function tells us the probability that $k$ out of the $n$ stars in that bin make it into the sample:
\begin{equation}
    \operatorname{Binomial}(k|n,q) = \binom{n}{k}q^k(1-q)^{n-k}.
\end{equation}
We will use the binomial probability distribution as our likelihood function $\operatorname{L}(q|k,n)=\operatorname{Binomial}(k|n,q)$ when modelling $q$. We note that $q$ is a probability on the unit interval, while most statistical machinery operates on quantities with the domain $(-\infty,+\infty)$. We avoid this issue by instead modelling the log-odds $x$
\begin{equation}
    x=\log\left(\frac{q}{1-q}\right) \Leftrightarrow q=\frac{1}{1+e^{-x}}.
\end{equation}
The log-likelihood is then, dropping constants and terms that only depend on $n$ and $k$,
\begin{equation}
    \log\operatorname{L}(x|k,n) = \left(k-\frac{n}{2}\right)x - n\log{\cosh{\left(\frac{x}{2}\right)}}, \label{eq:loglikelihood}
\end{equation}
and the derivative is
\begin{equation}
     \frac{\mathrm{d}\log\operatorname{L}}{\mathrm{d}x} = \left(k-\frac{n}{2}\right) - \frac{n}{2} \tanh{\left(\frac{x}{2}\right)}. \label{eq:loglikelihoodgrad}
\end{equation}
In Appendix \ref{sec:overflow} we give expressions for these two equations that are computationally efficient and avoid the risk of overflow.

In the next section we will develop a model that expresses each $x_{pmc}$ as a function of parameters $\mathbf{z}$ (the definition of the parameters $\mathbf{z}$ and their functional relationship with $x_{pmc}$ is deferred to that section) with those parameters being common across all bins and the functions being specific to each bin. It is the sharing of these parameters that causes our final estimated selection function to be correlated in position, magnitude and colour. The total log-likelihood of the parameters $\mathbf{z}$ is given by summing the log-likelihoods in each bin,
\begin{equation}
    \log\operatorname{L}_{\mathrm{total}}(\mathbf{z}) = \sum_{p,m,c}\log\operatorname{L}(x_{pmc}(\mathbf{z})|k_{pmc},n_{pmc}).
\end{equation}
We can optimise those parameters to maximise the inferred log-posterior (after assuming some prior on them) and thus obtain a best-estimate of the selection function.

For verification purposes, we can obtain an analytic estimate of the unknown selection function under the assumption that the selection probability $q_{pmc}$ does not depend on bin properties and is drawn from a $\operatorname{Beta}(\alpha,\beta)$ prior. The Beta probability density function is
\begin{equation}
    \operatorname{Beta}(x|\alpha,\beta) = \frac{\Gamma(\alpha+\beta)}{\Gamma(\alpha)\Gamma(\beta)}x^{\alpha-1}(1-x)^{\beta-1},
\end{equation}
defined for $x$ in the unit interval, where $\alpha>0$ and $\beta>0$ are parameters that determine the shape, and $\Gamma(y)$ is the Gamma function that generalises the factorial to real numbers. If our prior on $q_{pmc}$ is $\operatorname{Beta}(\alpha,\beta)$ then our posterior after observing that $k_{pmc}$ of the stars in the catalogue of $n_{pmc}$ stars are in the subset will be $\operatorname{Beta}(\alpha+k_{pmc},\beta+n_{pmc}-k_{pmc})$ (see \citetalias{PaperII} for an exposition). The mean of this posterior, \begin{equation}
    \frac{\alpha + k_{pmc}}{\alpha+\beta+n_{pmc}},
\end{equation}
then gives a convenient estimate of the selection probability. If we adopt a uniform prior on $q_{pmc}$ ($\alpha=\beta=1$), then this reduces to $(1+k_{pmc})/(2+n_{pmc})$. 

Virtually all previous selection functions evaluated for samples have simply quoted $k_{pmc}/n_{pmc}$ as the selection probability in the given bin \citep[e.g. ][]{Bovy2012Segue, Stonkute2016, Wojno2017, Chen2018, Mints2019, Rybizki2021}. $k_{pmc}/n_{pmc}$ is actually the mode of the posterior. For large $n_{pmc}$ this will approximately equate to the mean, however, for small $n_{pmc}$ they can take very different values. In the most extreme case, for bins with $n_{pmc}=0$ the mode becomes undefined which is understandable as the uniform prior has no mode.

By evaluating our model with the binomial likelihood function, our method is capable of estimating the same likelihood functions as all previous works listed above but correctly providing the expected value rather than the mode. Furthermore, we can fit far more general models in terms of position on the sky, apparent magnitude and colour as we'll demonstrate throughout the remainder of the paper.

\section{Model of the log-selection odds}
\label{sec:model}

We assume the log-odds that a star in the catalogue lying in position bin $p$, magnitude bin $m$ and colour bin $c$ is selected into the subset is a linear model,
\begin{equation}
    x_{pmc} = \mu_{pmc} + \sum_{i,j,k}B_{pmcijk}z_{ijk},
\end{equation}
where the $z_{ijk}$ are an arbitrary number of hidden variables for which we assume a $\operatorname{N}(0,1)$ prior, $B_{pmcijk}$ is the six-dimensional tensor that causes the selection log-odds to be correlated between $(p,m,c)$ bins, and $\mu_{pmc}$ is a three-dimensional tensor that gives an initial guess of the selection log-odds in each bin. The indices $i$, $j$ and $k$ are free to run over whichever values we wish. We stress that $x_{pmc}$ and $z_{ijk}$ are random variables whose values we determine by maximum posterior estimation with the binomial likelihood previously described, whilst $\mu_{pmc}$ and $B_{pmcijk}$ are constants we must choose in advance. We simplify our space of choices by assuming that the tensor $B_{pmcijk}$ is separable,
\begin{equation}
    B_{pmcijk} = P_{pi}M_{mj}C_{ck}.
\end{equation}
In the remainder of this section we will outline choices for these three tensors that lead to useful models of the selection log-odds.

\subsection{Position only}
\label{sec:position}
Suppose we only want to model the position dependency of the selection function $x_p$, which is equivalent to summing over the magnitude $m$ and colour $c$ indices and collapsing the corresponding hidden variable indices $j=k=1$. The star counts reduce to $k_p=\sum_{m,c}k_{pmc}$ and $n_p=\sum_{m,c}n_{pmc}$, the contribution of the magnitude and colour tensors reduces to a single number $\sigma$, and the mean must be at most one dimensional $\mu_p$,
\begin{equation}
    x_p = \mu_p + \sigma\sum_{i}P_{pi}z_i.
\end{equation}
We have placed no requirement on how the position indices were computed from the stellar positions: they could refer to pixels of a pixelisation scheme that covers the sky, or to fields of an astronomical survey that overlap, or even to membership of different star clusters. There are several useful choices for the tensor $\mathbfss{P}$.

\subsubsection{Trivial}
\label{sec:trivial}
The trivial choice is to assume that the $x_p$ are independent, which can be done by letting the indices $i$ run over the same values as $p$ and choosing $P_{pi}=\tau_i\delta_{pi}$ (which equals $\tau_p$ if $p=i$, zero otherwise). This causes the selection log-odds in each position bin to be independently and normally distributed, $x_p\sim\operatorname{N}(\mu_p,\sigma\,\tau_p)$, with the values $\mu_p$, $\sigma$ and $\tau_p$ being free parameters. This trivial choice should be used when there is no expectation that the selection function between position bins is correlated, which will usually be the case if the position bins represent disjoint fields or star clusters.

\subsubsection{Spherical harmonic}
\label{sec:sph_harmonics}
If the position bins have a meaningful average coordinate on the sky $(\theta_p,\varphi_p)$, then spherical harmonics - as the standard orthonormal basis on the sphere - are the obvious choice to define the tensor. The spherical harmonics are usually defined in terms of two indices $(\ell,m)$\footnote{In Sec. \ref{sec:position} $m$ will be used as a label for spherical harmonic modes, while in the rest of this paper it labels the magnitude bins.} rather than one, altering the expression for $x_p$,
\begin{equation}
    x_p = \mu_p+\sigma\sum_{\ell=0}^{\ell_{\mathrm{max}}}\sum_{m=-\ell}^{m}P_{p\ell m}z_{\ell m}, \label{eq:sphericalharmonic}
\end{equation}
although we note that the two indices can be stacked to instead run over one index. Similarly, the hidden variables satisfy $z_{\ell m}\sim\operatorname{Normal}(0,1)$. The tensor is defined by $P_{p\ell m} = \tau_{\ell m}Y_{\ell m}(\theta_p,\varphi_p)$ using the real spherical harmonics, which are obtained from the standard complex spherical harmonics $Y_\ell^m$ through
\begin{equation}
    Y_{\ell m} =
\begin{cases}
\displaystyle \frac{i}{\sqrt{2}} \left(Y_\ell^m - (-1)^m\, Y_\ell^{-m}\right) & \text{if}\ m<0\\
\displaystyle  Y_\ell^0 & \text{if}\ m=0\\
\displaystyle  \frac{1}{\sqrt{2}} \left(Y_\ell^{-m} + (-1)^m\, Y_\ell^{m}\right) & \text{if}\ m>0.
\end{cases}
\end{equation}
The complex spherical harmonics are themselves defined by
\begin{equation}
    Y_{\ell}^{m}(\theta,\varphi) = \sqrt{\frac{2m+1}{4\pi}\frac{(\ell-m)!}{(\ell+m)!}}e^{im\varphi}\operatorname{P}_{\ell}^{m}(\cos{\theta}),
\end{equation}
where $\operatorname{P}_{\ell}^{m}(x)$ is the associated Legendre polynomial. The parameters $\tau_{\ell m}$ scale the hidden variables $z_{\ell m}$, allowing us to, for instance, prevent the model from over-fitting small angular features by having $\tau_{\ell m}$ decay with increasing  $\ell$. The $\tau_{\ell m}$ could alternatively be given values that capture the expected amount of structure in the selection function at different angular scales. For instance, if we have an angular power spectrum $C_{\ell}$ then we could choose $\tau_{\ell m}=\sqrt{C_\ell}$, which we justify in Appendix \ref{sec:powerspectrum}. However, in practise, we find that setting $\tau_{\ell m}=1$ gives adequate performance. The natural choice for the means $\mu_{p}$ is to set them equal to zero, unless there is prior information supporting a different mean for specific pixels. During optimisation the coefficient $z_{00}$ of the spherical harmonic mode that is equal-valued across the sky will adjust the mean across the entire sky.

The spherical harmonic basis only requires that the position bins have a meaningful average coordinate, and so could be used with disjoint fields. However, a common use case will have the entire sky be tiled by position bins, thus requiring Eq. \ref{eq:sphericalharmonic} to be evaluated a large number of times with a computational cost that is $\mathcal{O}(\ell_{\mathrm{max}}^2N_p)$, where $N_p$ is the number of position bins. We can minimise that computational cost if we use the HEALPix \citep{Gorski2005} pixelisation scheme, which was designed with spherical harmonic calculations in mind. The HEALPix schemes places pixel centres equidistantly along isolatitude rings and \citet{Reinecke2011} showed how this structure can be exploited to reduce the computational complexity to $\mathcal{O}(\ell_{\mathrm{max}}^3)$. It is generally the case that $\ell_{\mathrm{max}}\ll N_p$ making this a significant saving.

The weakness of the spherical harmonic model is that the modes are not localised on the sky, and thus a change to the coefficient multiplying just one of the modes changes the selection function everywhere. If there is a localised region where the selection significantly deviates from the mean and is well-populated by the intermediate sample (for instance, if all the stars in a globular cluster did not make it into the sample) then the model will need to use a large number of high-order modes to both capture the local deviation and to cancel out the effect of those modes elsewhere. If $\ell_{\mathrm{max}}$ is set too low then the cancellation will not be exact, with the large number of stars in the local deviation leaving residual `ripples' that cause the selection function to be poorly reproduced in nearby parts of the sky.

\subsubsection{Needlets}
\label{sec:sph_needlets}
Needlets, sometimes called spherical wavelets, are an alternative to spherical harmonics that are localised on the sphere. They are defined in terms of the spherical harmonics,
\begin{align}
    \psi&_{jk}(\theta,\varphi) \nonumber \\
    &= \sqrt{\lambda_{jk}} \sum_{\ell=0}^{\ell_{\mathrm{max}}}\operatorname{b}_{\ell}(j)\sum_{m=-\ell}^{\ell}\bar{Y}_{\ell}^{m}(\theta,\varphi)Y_{\ell}^{m}(\theta_{jk},\varphi_{jk}),
    \label{eq:psi}
\end{align}
where the $(\theta_{jk},\varphi_{jk})$\footnote{In Sec. \ref{sec:position} $j$ and $k$ will be used as labels for spherical needlet modes, while in the rest of this paper they label the magnitude and colour hidden variables.} are the pixel centres of order $j$ HEALPix grids, $\lambda_{jk}=4\pi/N_j$ is the area taken up by one of the $N_j$ pixels in those grids, and $\operatorname{b}_{\ell}(j)$ is a window function that weights the contribution of each spherical harmonic to the needlets at order $j$. The spherical harmonic addition formula allows us to simplify this expression,
\begin{align}
    \psi&_{jk}(\theta,\varphi) \nonumber \\
    &= \sqrt{\lambda_{jk}} \sum_{\ell=0}^{\ell_{\mathrm{max}}}\operatorname{b}_{\ell}(j) \frac{2\ell+1}{4\pi}\operatorname{P}_{\ell}\left(\cos{(\phi_{jk}(\theta,\varphi))}\right),
\end{align}
allowing us to compute the needlets directly using the standard Legendre polynomials $\operatorname{P}_{\ell}(x)$, where $\phi_{jk}(\theta,\varphi)$ is the angular great circle distance between $(\theta,\varphi)$ and $(\theta_{jk},\varphi_{jk})$.

There are a number of common choices for the window function $\operatorname{b}_{\ell}(j)$, however the obvious choice for modelling selection functions is that suggested by \citet{Geller2009} and popularised in astrophysics by \citet{Scodeller2011}:
\begin{equation}
    \operatorname{b}_{\ell}(j|B,\nu) = \left(\frac{\ell}{B^j}\right)^{2\nu}e^{-\frac{\ell^2}{B^{2j}}},
\end{equation}
where $B,\nu>1$ parameterise a family of possible window functions. Needlets defined by these window functions have extremely good localisation on the sphere - at fixed angular distance $\phi$ their tails decay like $\exp{(-B^{2j}\phi^2/4)}$ with increasing $j$. Needlets with this window function are termed `Mexican' needlets in the literature, however we prefer to term them `Chi-square' needlets, because the weighting function resembles the probability density function of a Chi-square random variable with $\nu$ degrees of freedom.

\begin{figure}
    \centering
	\includegraphics[width=\linewidth]{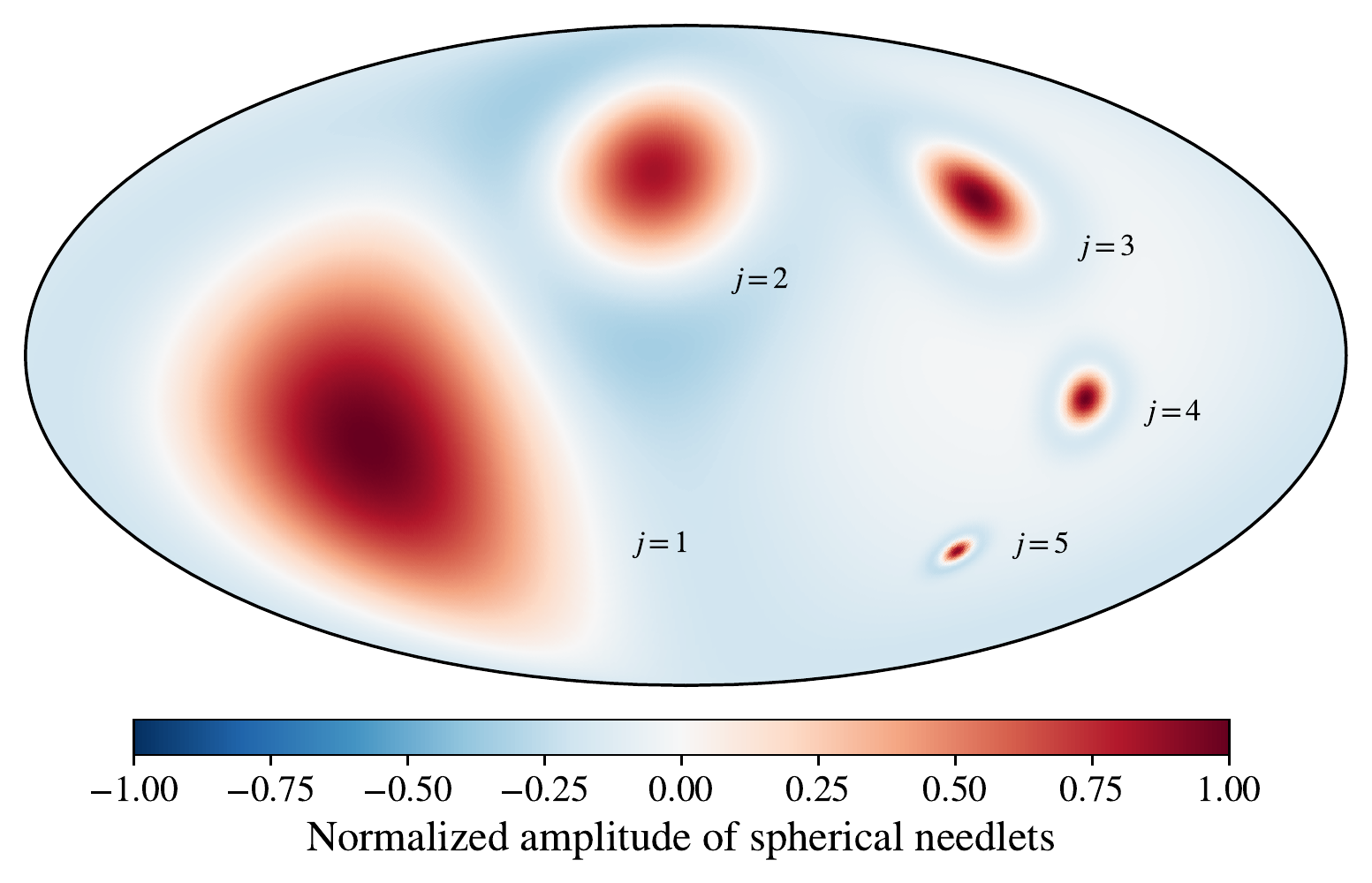}
	\caption{Spherical needlets are weighted sums of spherical harmonics across a range of angular modes that result in a spatially-localised basis for functions on the sphere. Shown are spherical needlets of different orders $j$, with higher orders having smaller angular scales. The needlets have each been normalised to have their peak at one.}
	\label{fig:needlets}
\end{figure}

The expression for $x_p$ in terms of the needlets is
\begin{equation}
    x_p = \mu_p + \sigma\sum_{j=0}^{j_{\mathrm{max}}}\sum_{k=0}^{N_j}P_{pjk}z_{jk}, \label{eq:needlet}
\end{equation}
with $P_{pjk}=\tau_{jk}\psi_{jk}(\theta_p,\varphi_p)$. The parameters $\tau_{jk}$ serve a similar purpose to the $\tau_{\ell m}$ variables in the spherical harmonic case, allowing us to constrain the freedom of the model at different angular scales. This expression can be computationally expensive to evaluate when the position bins are themselves on a HEALPix grid - for instance, if the position bins are on an order 7 HEALPix grid and $j_{\mathrm{max}}=6$, then $P_{pjk}$ will have 12,884,115,456 elements. Fortunately, almost all of these elements are zero due to the excellent localisation properties of Chi-square needlets, and we can save on computations and memory by exploiting that sparsity. There is no equivalent to the `zeroth mode' of the spherical harmonic model and so we add on a `$j=-1$' mode which is equal-valued across the sky. We illustrate first, second, third, fourth and fifth order Chi-square needlets with $B=2$ and $\nu=1$ in Fig. \ref{fig:needlets}, where we have normalised each needlet to have its maximum at one. Note that each needlet falls to a negative value then decays to zero with increasing angular distance from the needlet centre.

\subsection{Magnitude and colour}
Suppose we only want to model the magnitude dependency of the selection function $x_m$, which is equivalent to summing over the position $p$ and colour $c$ indices and collapsing the corresponding indices $i=k=1$. The star counts reduce to $k_m=\sum_{p,c}k_{pmc}$ and $n_m=\sum_{p,c}n_{pmc}$, the contribution of the position and colour tensor reduces to a single number $\sigma$, and the mean must be at most one dimensional $\mu_m$,
\begin{equation}
    x_m = \mu_m + \sigma\sum_{j}M_{mj}z_j.
\end{equation}
We label the magnitude at the centre of each bin $y_m$ and assume that $j$ runs over the same values as $m$. This formulation is identically equivalent to assuming that the $x_m$ follow a Multivariate Normal distribution with mean vector $\bm{\mu}$ and covariance matrix $\sigma^2\mathbfss{M}\mathbfss{M}^{\intercal}$. This is usually referred to as a Gaussian Process. This correspondence suggests that we should proceed from the other direction, by choosing a Gaussian process kernel function $K(y,y')$ from the list of common examples in Tab. \ref{tab:kernels}, evaluating the implied covariance matrix $K_{mj}=K(y_m,y_j)$, computing the Cholesky decomposition $L_{mj}$ satisfying $\mathbfss{L}\mathbfss{L}^{\intercal}=\mathbfss{K}$, then setting $M_{mj} = L_{mj}$. This has the advantage of allowing us to reason about the covariance we want to capture on different magnitude scales and permitting us to use the infinite variety of Gaussian Process kernels in the literature to realise those covariances. Which kernel to use will depend on the use case and the user's prior expectation of the smoothness and lengthscale of the selection function being estimated, but a good starting point is to assume a Squared Exponential kernel with a variance of one and a lengthscale equal to the bin-width in the magnitude dimension. Readers wishing to gain a greater understanding of the large family of kernels commonly in use may find Chapter~4 of \citet{rasmussen2006} enlightening.

\begin{figure}
    \centering
	\includegraphics[width=\linewidth]{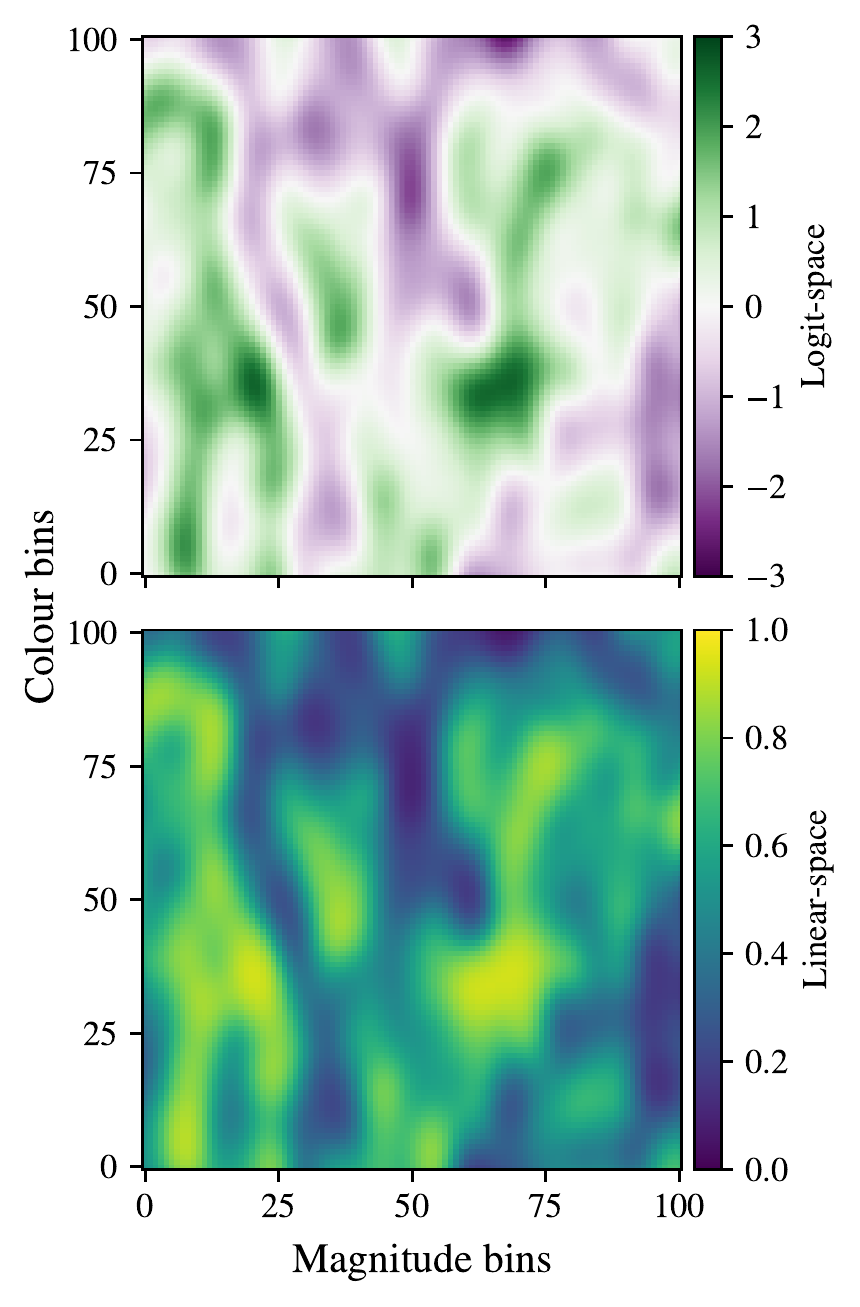}
	\caption{Gaussian process across two dimensions provide a non-parametric way to model a selection function across magnitude and colour. In the top panel we show a draw from a Gaussian process and in the bottom panel we transform that draw into the unit interval using the logit function to allow it to be interpreted as a selection probability.}
	\label{fig:gaussianprocess}
\end{figure}

We can apply this same logic to infer that the colour tensor should be the Cholesky decomposition of a covariance matrix derived from a Gaussian Process kernel evaluated at the colour bin centres. A model of the selection log-odds over magnitude and colour only would have the form
\begin{equation}
    x_{mc} = \mu_{mc} + \sigma\sum_{j,k} M_{mj}C_{ck}z_{jk}.
\end{equation}
Note that this is identically equivalent to a two-dimensional Gaussian Process where the common assumption has been made that the covariance across the two dimensions is separable. We illustrate one random draw from a two-dimensional Gaussian Process in the top panel of Fig. \ref{fig:gaussianprocess}, where in the magnitude dimension we have used a Squared Exponential kernel with a variance of one and a lengthscale of five, while in the colour dimension we have used a Rational Quadratic kernel with a variance of one, a lengthscale of ten and a power law of one. The bottom panel of Fig. \ref{fig:gaussianprocess} shows the selection function obtained when we compute $1/(1+e^{-x_{mc}})$.

\subsection{Position, magnitude and colour}
Combining all of the ingredients above, we arrive at a highly general linear model for a selection function across position, magnitude and colour. We assume that the selection log-odds are a linear sum of independent Normal random variables $z_{ijk}\sim\mathrm{Normal}(0,1)$,
\begin{equation}
    x_{pmc} = \mu_{pmc} + \sum_{i,j,k} P_{pi}M_{mj}C_{ck}z_{ijk},
\end{equation}
where the position tensor is given by either the trivial, spherical harmonic or spherical needlet basis for functions on the sphere, and the magnitude and colour tensor are the Cholesky decomposition of a covariance matrix derived from Gaussian Process kernels evaluated at the magnitude and colour bin centres.

\section{Selection function toolbox}
\label{sec:toolbox}
We have implemented the model described above in a new open-source \textsc{Python} package \textsc{selectionfunctiontoolbox}\footnote{\url{https://github.com/gaiaverse/selectionfunctiontoolbox}}. There are three main tools in our toolbox which each use a different set of modes to model the dependence of the selection function on a star's position:  the \textit{wrench} (positions are trivially independent - see \ref{sec:trivial} - colour-magnitude selection functions need to be `bolted' together to form an all-sky selection function), the \textit{hammer} (spherical harmonics - see \ref{sec:sph_harmonics} - each mode causes ringing around the entire sphere), and the \textit{chisel} (spherical needlets - see \ref{sec:sph_needlets} - modes are spatially-localised and so the model can make localised changes by altering a single parameter).

Each model in \textsc{selectionfunctiontoolbox} is written in the \textsc{stan} probabilistic programming language \citep{Carpenter2017} accessed through the \textsc{cmdstanpy} Python package, which we optimise using the Limited-memory BFGS \citet{Liu1989} algorithm. \textsc{stan} is an automatically-differentiable language obviating the need for us to write the derivatives of our model, but this has the downside of excessive computation and memory requirements if several large matrices are multiplied. We have implemented two different approaches to minimise the computations required when multiplying the magnitude or colour Cholesky matrices, covering the two limiting cases of the magnitude or colour random variables being perfectly uncorrelated or perfectly correlated. If the random variables are only weakly correlated then the Cholesky matrix will be close to diagonal, meaning that we can use a sparse representation to avoid multiplying (and propagating gradients related to) a large number of zeros. If the random variables are highly correlated then we can compute a pivoted, lower dimensional approximation to the Cholesky matrix that will incur a lower multiplication cost. We utilise the sparse matrix functionality in \textsc{scipy} and an open-source implementation\footnote{\url{https://github.com/NathanWycoff/PivotedCholesky}} of the low-rank pivoted Cholesky decomposition \citep{Harbrecht2012}. 

To illustrate the use of \textsc{selectionfunctiontoolbox}, we use it to deduce the selection function of the Sloan Digital Sky Survey's \citep{Blanton2017} Apache Point Observatory Galactic Evolution Experiment (APOGEE, \citealp{Majewski2017,Nidever2015,Wilson2019}), specifically the main red giant survey in the sixteenth data release \citep{Ahumada2020}. We specifically select only those stars in the main red giant sample, defined by \textsc{EXTRATARG==0}. APOGEE selected stars from the Two Micron All-Sky Survey (2MASS, \citealp{Skrutskie2006}) that lay in specific circular fields across the sky and that satisfied cuts in $H$ magnitude and dereddened $J-K$ colour, with those magnitude-colour cuts varying between fields. If APOGEE had observed all of the stars satisfying these cuts in position, magnitude and colour, then the selection function would be binary: unity if a star satisfies the cuts, zero otherwise. However, the finite width of the fibres in APOGEE's multi-object fibre spectrograph limited the number of stars they could observe simultaneously in a given field, causing the selection function to be less than one even for stars satisfying the cuts.

\begin{figure}
    \centering
	\includegraphics[width=\linewidth]{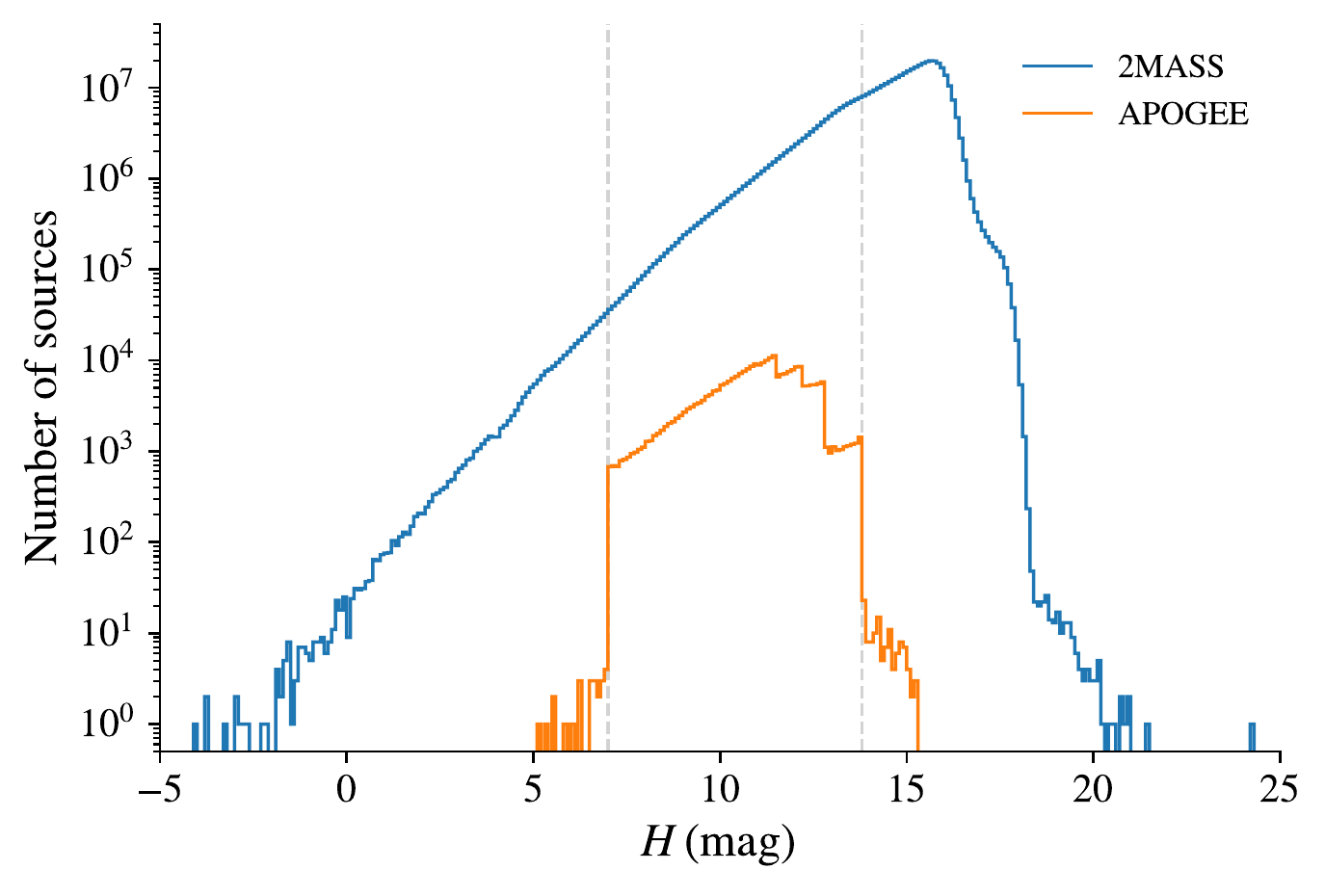}
	a) Apparent magnitude histogram
	\includegraphics[width=\linewidth]{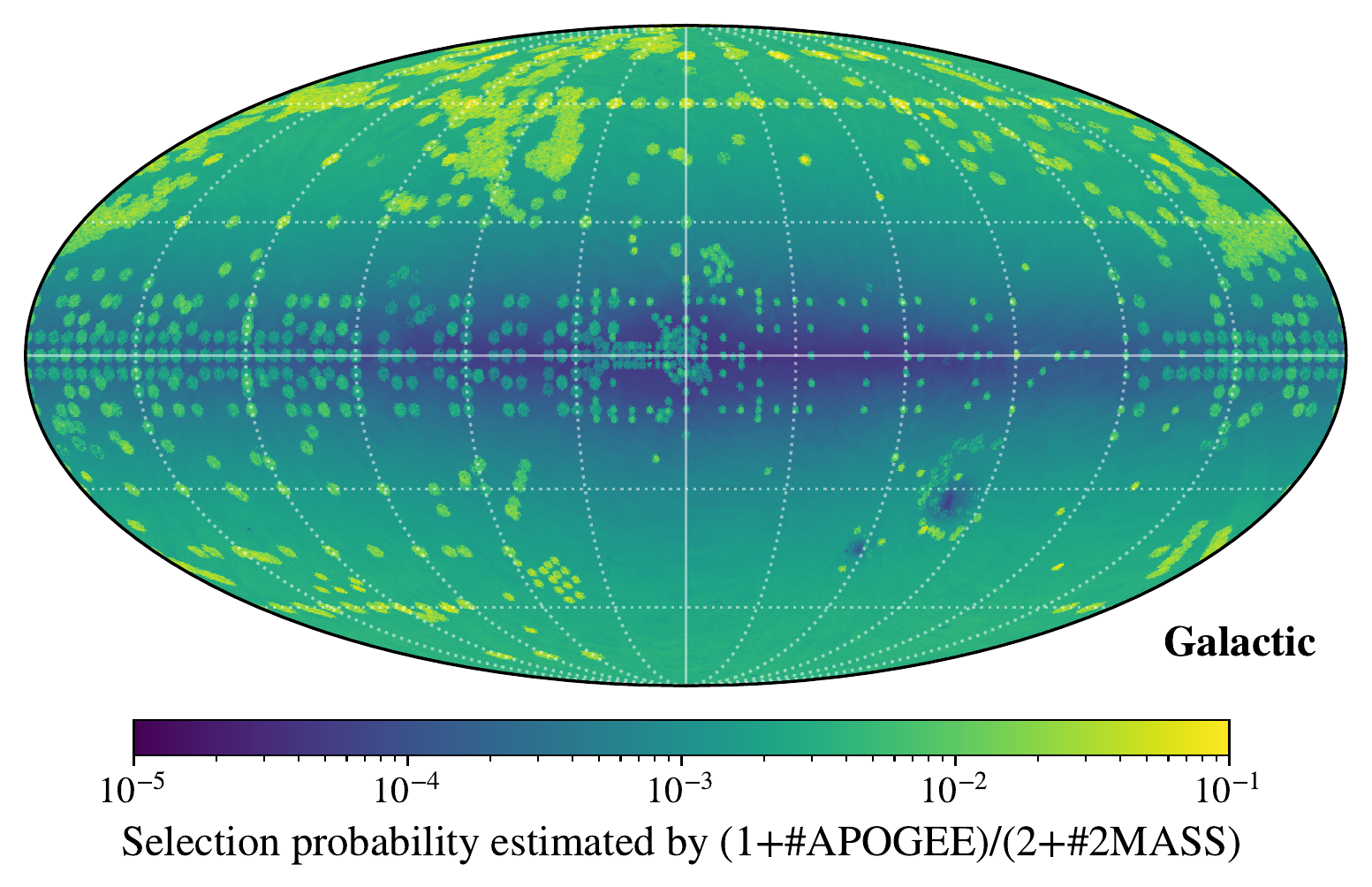}
	b) Estimated selection probability
	\caption{\textbf{Top:} Histogram of the number of sources in 2MASS and the APOGEE DR16 main red giant survey as a function of $H$ magnitude. The grey dashed lines denote the nominal $7<H<13.8\;\mathrm{mag}$ limits of the APOGEE survey selection. \textbf{Bottom:} Estimate of the probability that a source in 2MASS is in the APOGEE DR16 main red giant survey in pixels on the sky.}
	\label{fig:apogeetwomass}
\end{figure}

In the top panel of Fig. \ref{fig:apogeetwomass} we show the apparent magnitude distribution of the sources in 2MASS and the APOGEE DR16 main red giant survey in $0.1\;\mathrm{mag}$ bins, illustrating that the stars observed by APOGEE are a small subset of those reported in 2MASS. By counting the number of sources in APOGEE $k$ and 2MASS $n$ in each $\textsc{nside}=128$ HEALPix pixel, we can estimate the typical selection probability in that pixel as $(1+k)/(2+n)$ and we show this quantity across the sky in the bottom panel of Fig. \ref{fig:apogeetwomass}. As discussed in Sec. \ref{sec:logodds}, this quantity is the mean of your posterior on the bias of a coin after flipping a coin $n$ times and observing $k$ heads, starting from a uniform prior on the coin's bias. This figure shows that - even for stars lying in one of the APOGEE fields - the probability of being observed was greatly reduced in fields near the Galactic plane with high stellar number density. That our rough estimate of the selection probability varies outside of the APOGEE fields reflects that the more 2MASS sources there are in a pixel that weren't observed by APOGEE, the more confident you can be that the selection probability was very small. This estimate of the selection probability ignores any magnitude dependence and is highly sensitive to the number and shape of the pixels (in the limit where every star lies in a unique pixel, the estimate will either be $1/2$ if there are no stars in that pixel, $1/3$ if there is a star that wasn't observed by APOGEE, or $2/3$ if there is a star that was observed by APOGEE), motivating us to construct more reliable estimates of the selection probability in terms of magnitude and position using the three tools in \textsc{selectionfunctiontoolbox}. All of the following examples were generated using a \textsc{Jupyter} notebook that is available on-line\footnote{\url{https://github.com/gaiaverse/selectionfunctiontoolbox/blob/main/Examples/Example.ipynb}} and we encourage readers to try it out. To avoid the computations being overly demanding we have necessarily lowered the magnitude and position resolution of our selection functions such that they each take less than one hour to estimate on a single core.

\subsection{Model A}
Our first attempt at modelling the selection function assumes that we have no a priori knowledge of the dependence on magnitude and position - we simply have a catalogue and a subset drawn from that catalogue. In this situation our intermediate sample is the entire 2MASS catalogue, while our science sample comprises the stars in the APOGEE DR16 red giant sample. We count the stars in magnitude and position bins in APOGEE $k_{pm}$ and 2MASS $n_{pm}$, using $0.5\;\mathrm{mag}$ bins over $-5<H<25\;\mathrm{mag}$ and an $\textsc{nside}=32$ HEALPix grid across the sky. We assume that the mean of the model is $\mu=-10$, that the magnitude dependence is described by a Rational Quadratic kernel with a variance of $1$, a lengthscale of $1\;\mathrm{mag}$ and a powerlaw of $1$, and use a spherical harmonic basis up to $l=64$ to model the position dependence (the `hammer').

\begin{figure}
    \centering
	\includegraphics[width=\linewidth]{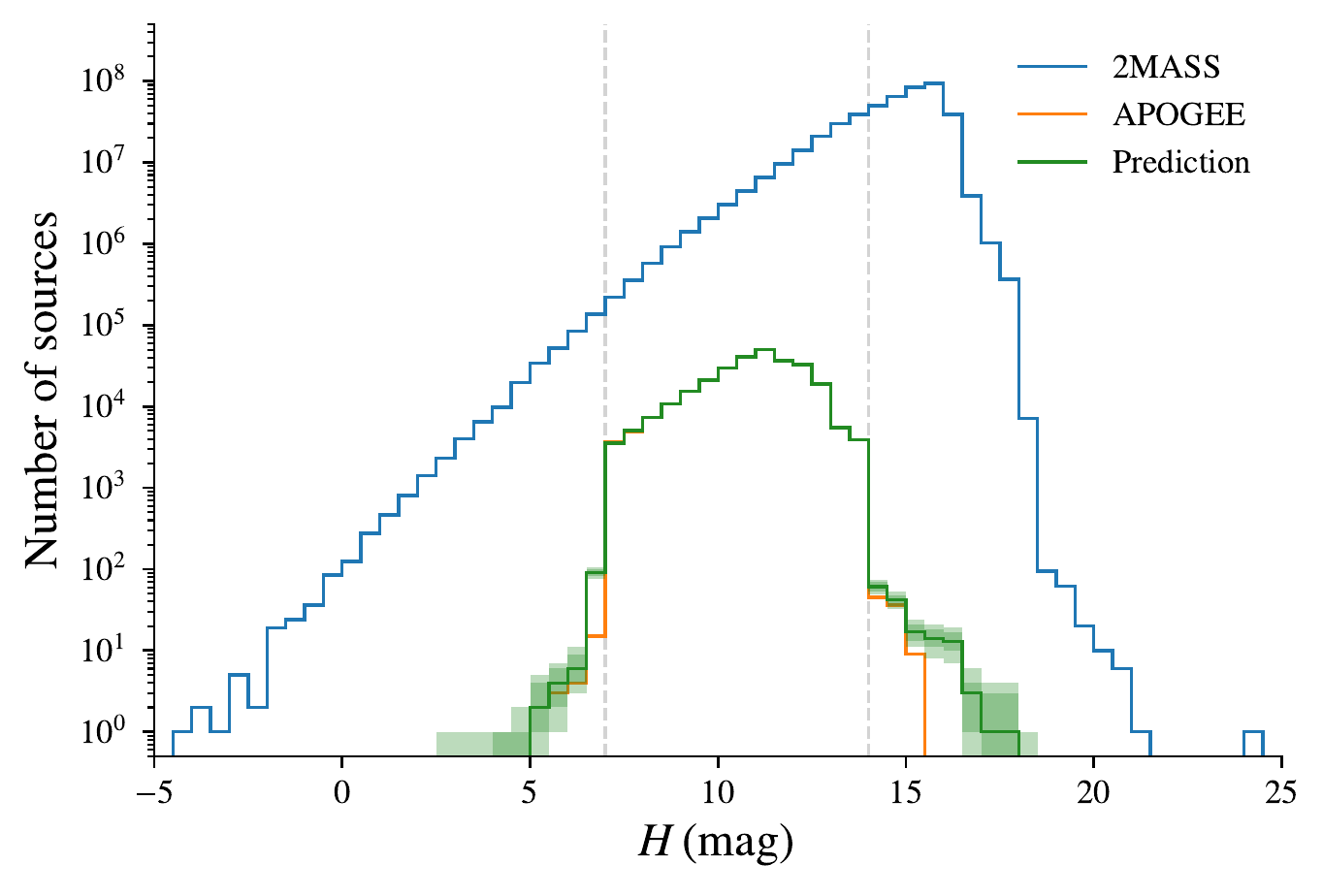}
	\caption{The numbers of 2MASS (blue) and APOGEE (orange) stars in each magnitude bin, and the number of APOGEE stars predicted by our Model A (green). We show the one-sigma and two-sigma confidence intervals of our prediction as shaded regions. The grey dashed lines enclose the interval $7 < H < 14\;\mathrm{mag}$ that contains almost all of the APOGEE stars.}
	\label{fig:naive_magnitude_selection}
\end{figure}

In Fig. \ref{fig:naive_magnitude_selection} we show the total number of 2MASS and APOGEE stars in each $0.5\;\mathrm{mag}$ bin, as well as a prediction for the number of APOGEE stars given the stars in 2MASS and our inferred selection function. The prediction is calculated by assuming that the number of APOGEE stars in each magnitude bin is Poisson-distributed, with a rate $\lambda_m$ given by summing the number of stars in 2MASS multiplied by our inferred selection function across position,
\begin{equation}
    \lambda_m = \sum_{p}\frac{n_{pm}}{1+e^{-x_{pm}}}.
\end{equation}
The shaded regions indicate the one-sigma and two-sigma confidence intervals on the predicted number of APOGEE stars. Note that this Poisson-formulation is only an approximation. The predicted number of APOGEE stars in each position-magnitude bin is binomially-distributed and the sum of these across position bins is thus Poisson binomial distributed\footnote{The Poisson binomial probability distribution describes the probability of the number of heads when $n$ coins with different biases are flipped. It requires summing over the probability of each combination of coins which could have resulted in a given number of heads, and thus is computationally-demanding when $n$ is large.}, but Poisson binomial random variables can be approximated by Poisson random variables. The predicted number of APOGEE stars in each magnitude bin closely matches the observed number within the interval $7 < H < 14\;\mathrm{mag}$ that contains almost all of the APOGEE stars. Our inferred selection function overestimates the number of APOGEE stars just below $H=7\;\mathrm{mag}$ due to our assumption that the selection function is correlated over lengthscales of $1\;\mathrm{mag}$, which similarly causes the slight overestimate at the faint end.

\begin{figure*}
    \centering
	\includegraphics[width=\linewidth]{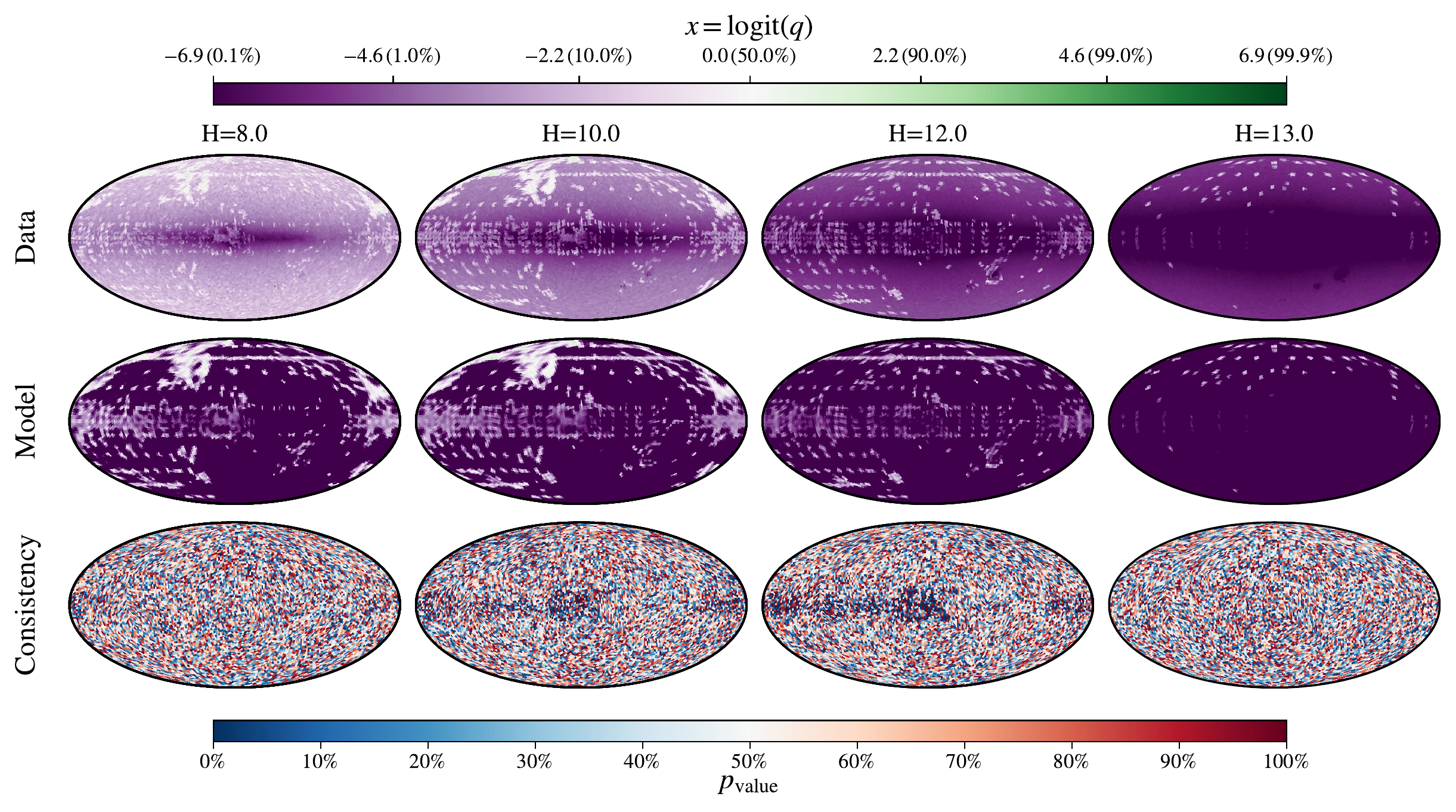}
	\caption{\textbf{Top:} A map of one plus the number of APOGEE stars in each pixel divided by two plus the number of 2MASS stars, which provides a simple estimate of the probability that stars in 2MASS at the labelled magnitude were selected in APOGEE. \textbf{Middle:} Our inferred selection probability under Model A. \textbf{Bottom:} A p-value test of the consistency of our model with the data - $p_{\mathrm{value}}$ should be uniform between 0\% and 100\% if our model is well-fit, clustered at either end if it is under-fit, or clustered at 50\% if it is over-fit.}
	\label{fig:naive_position_selection}
\end{figure*}

Along the top row of Fig. \ref{fig:naive_position_selection} we show a simple estimate of the selection function $(1+k_{pm})/(2+n_{pm})$ at different magnitudes. We stress that the gradient in colour outside the APOGEE fields reflects that this simple estimator can only return small selection probabilities in regions with many 2MASS stars. Along the middle row we show our inferred selection probability which does not share this weakness, because the selection function is correlated across the sky. The bottom row shows the p-value of the data in each pixel given our model and thus quantifies the consistency between our inferred selection probability and the data. The quantity $p_{\mathrm{value}}$ should be uniformly random if our model is well-fit to the data, will be clustered at 0\% or 100\% if we have under-fit to the data, and clustered at an intermediate value if we have over-fit the data. See the companion paper for a more detailed exposition on this quantity. Overall we have achieved a good fit to the data, except in the Galactic plane. This occurs because the spatial model is having to compromise fitting the relatively high selection probability within the fields against matching the small selection probability outside them - a contrast which is greatest in the Galactic plane because the density of 2MASS sources allows the model to be highly confident that the selection probability is very small outside the APOGEE fields.

\subsection{Model B}

Our second attempt at modelling the selection function resolves the issues of Model A by incorporating knowledge of the APOGEE targeting. We know that sources were only considered for incorporation in APOGEE if their magnitude lay in $7 < H < 13.8\;\mathrm{mag}$ (ignoring the small number of targets outside this range) and if they were inside one of the APOGEE fields. The issues with Model A in both magnitude and position space were due to the selection probability needing to compromise estimating the selection probability of stars within these regions whilst returning zero selection probability outside them, which caused the selection probability to be badly estimated on the borders of those regions. 

We can improve our estimate of the selection function by dividing it into a known component (which is one if a star satisfies these requirements and zero otherwise) and an unknown component (which accounts for the remainder of the selection effects). In the case of APOGEE, the known component could include knowledge of whether the star falls on one of the APOGEE fields, while the unknown component will be attempting to model random processes such as whether a star satisfying all the known selections cuts was randomly allocated a spectroscopic fibre. We can estimate this unknown component by comparing APOGEE to the intermediate sample drawn from 2MASS which satisfies the position and magnitude requirements. The advantage of doing this is that the model does not need to capture the sharp changes in selection probability that occur on the boundary of the magnitude selection or on the edges of fields.

We select all stars in APOGEE and 2MASS that satisfy $7 < H < 13.8\;\mathrm{mag}$ and that lie inside one of the APOGEE fields. We count the stars in magnitude and position bins in APOGEE $k_{pm}$ and 2MASS $n_{pm}$, using $0.4\;\mathrm{mag}$ bins over $7<H<13.8\;\mathrm{mag}$ and an $\textsc{nside}=32$ HEALPix grid across the sky. We assume that the mean of the model is $\mu=-10$, that the magnitude dependence is described by a Rational Quadratic kernel with a variance of $\sigma^2=1$, a lengthscale of $1\;\mathrm{mag}$ and a powerlaw of $\alpha=1$, and use a spherical needlets basis to model the position dependence (the `Chisel') with orders up to and including $j=5$ and use a Chi-Squared weighting function with $B=2$ and $p=1$.

\begin{figure}
    \centering
	\includegraphics[width=\linewidth]{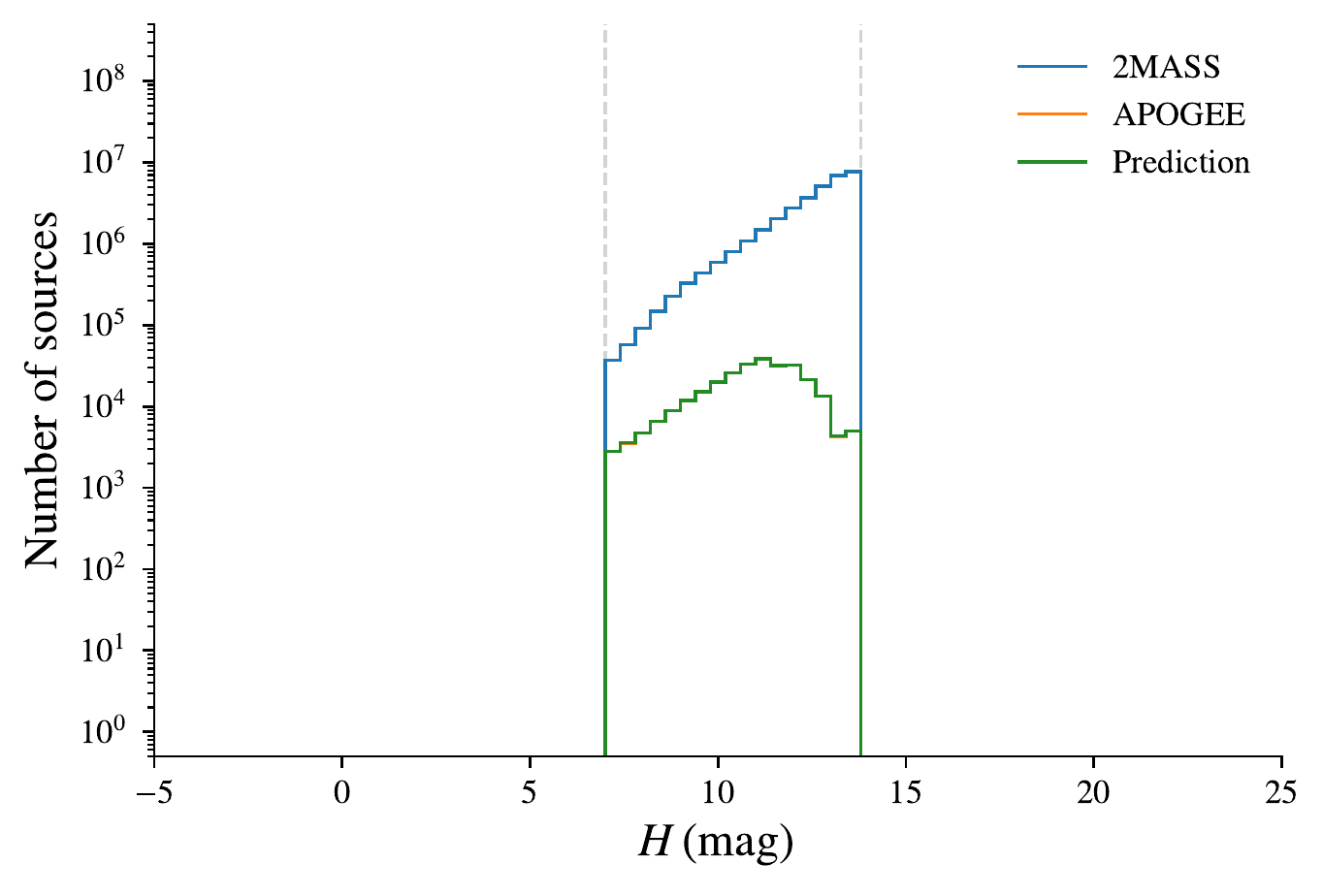}
	\caption{Same as Fig. \ref{fig:naive_magnitude_selection} but for Model B. The right grey line has been moved to $H=13.8\;\mathrm{mag}$, which is the faintest limit of the sample.}
	\label{fig:parent_magnitude_selection}
\end{figure}

\begin{figure*}
    \centering
	\includegraphics[width=\linewidth]{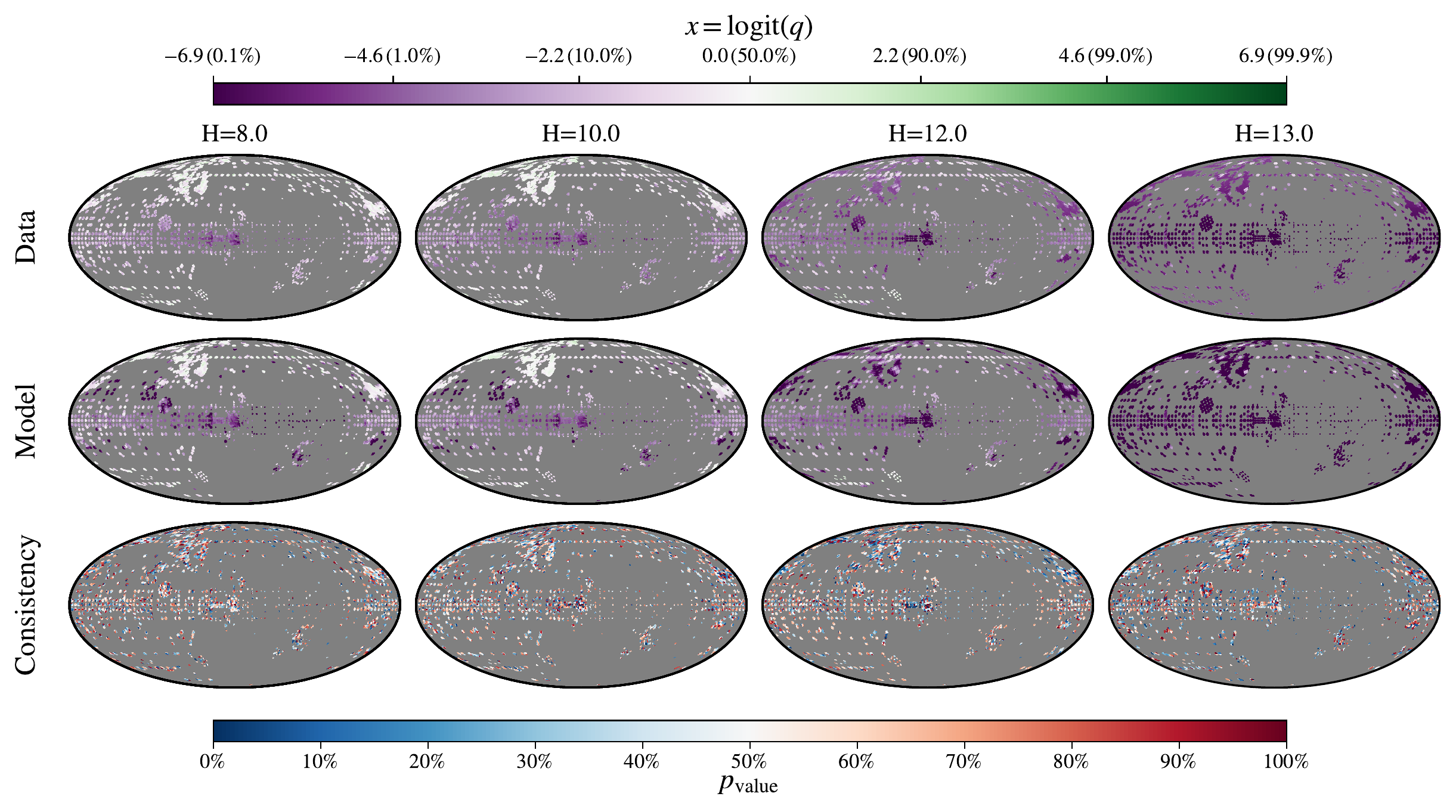}
	\caption{Same as Fig. \ref{fig:naive_position_selection} but for Model B. A mask has been applied to grey out regions outside APOGEE fields.}
	\label{fig:parent_position_selection}
\end{figure*}

In Fig. \ref{fig:parent_magnitude_selection} we show the total number of 2MASS and APOGEE stars in each $0.4\;\mathrm{mag}$ bin, as well as a prediction for the number of APOGEE stars given the stars in 2MASS and our inferred selection function. We perfectly recover the the magnitude distribution of the APOGEE sources. In Fig. \ref{fig:parent_position_selection} we show the performance of our model across the sky. A key point to note is that Model B does fit the selection function in bins outside of any fields, but these fields have $n_{pm}=k_{pm}=0$ and thus exert no influence on the fit inside the fields. We have applied a mask to every panel of Fig. \ref{fig:parent_position_selection} to prevent the figure being visually-dominated by the redundant parts of the model.

\begin{figure}
    \centering
	\includegraphics[width=\linewidth]{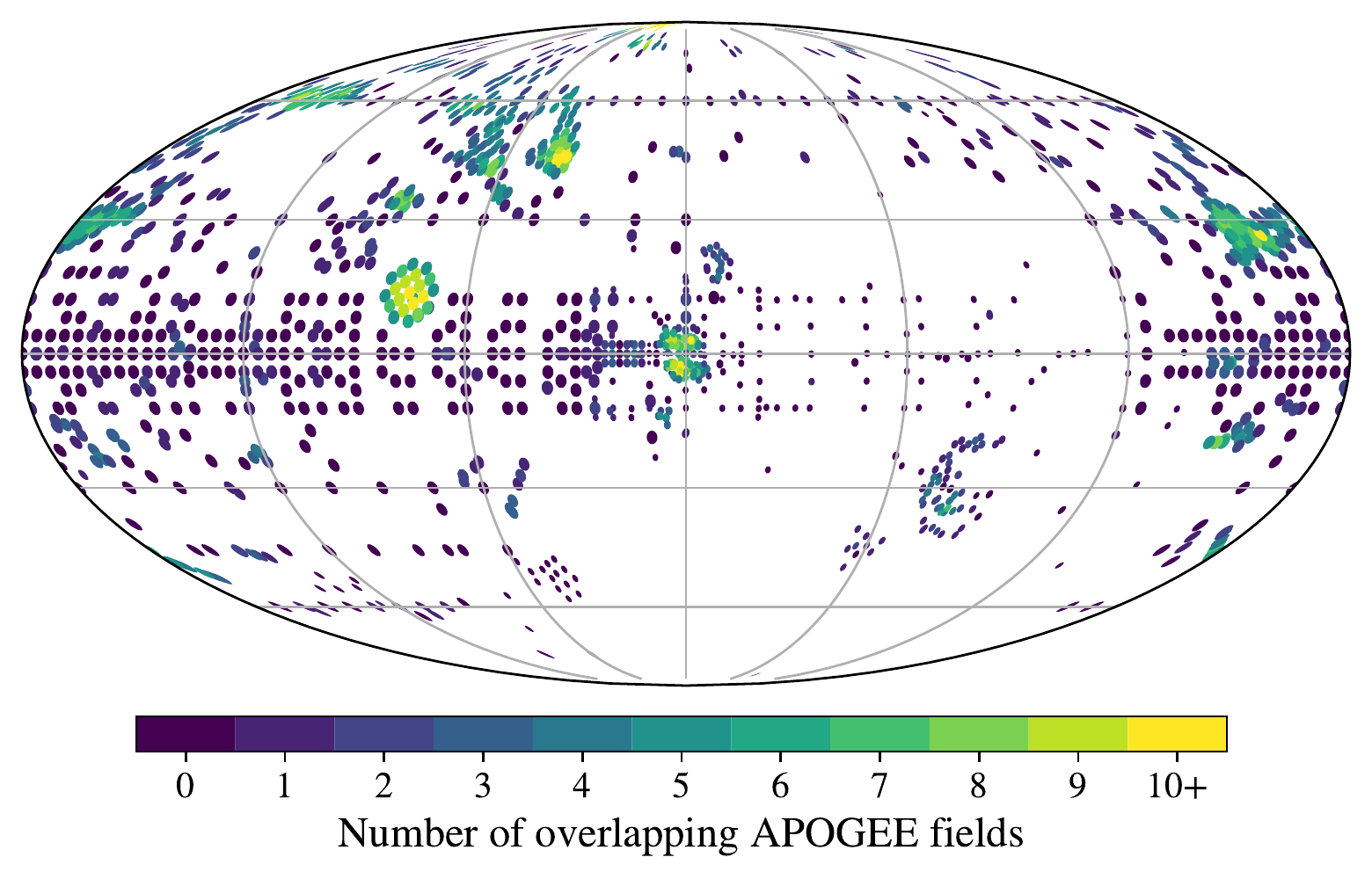}
	\caption{Each APOGEE field is plotted as a circle on a Mollweide projection in Galactic coordinates, and coloured by the number of other APOGEE fields that it overlaps with.}
	\label{fig:apogee_overlap}
\end{figure}

The bottom row of Fig. \ref{fig:parent_position_selection} shows that the consistency of our inferred selection function with the data is generally good, however there are three regions where there is some evidence of under- or over-fitting: the Galactic centre, the cluster of fields above the equator at the right hand edge, and the cluster of fields in the upper left quadrant. These issues are symptomatic of a structural issue with Model B - it does not incorporate our knowledge that the selection function increases where fields overlap. In Fig. \ref{fig:apogee_overlap} we show the location of APOGEE field as a circle, coloured by the number of other APOGEE fields that it overlaps with. The areas of the sky where fields most overlap coincide with the regions where Model B is most inconsistent with the data. A star lying in the overlap between two fields has two opportunities to be observed by APOGEE, and thus the selection function should be greater in these overlaps. However, these overlaps between fields that are a few degrees across are not resolved by the $1.8^{\circ}$-wide pixels in our $\textsc{nside}=32$ HEALPix bins, causing Model B to have to compromise between fitting stars inside and outside of these overlaps.

\subsection{Model C}

Our third attempt at modelling the selection function attempts to resolve the issues of Model B by incorporating knowledge that APOGEE fields overlap. We make the assumption that the observations made in each field were independent of observations made in other fields, and further that the magnitude selection was the same across the entirety of each field. This allows us to define each APOGEE field to be a unique position `bin' and to model the magnitude selection in each of these position bins as independent (the `Wrench'). The reason that we have termed this tool in \textsc{selectionfunctiontoolbox} the `Wrench' is that it is incumbent on the user to `bolt' the selection functions for each field together into one selection function applicable to the entire sky. For instance, if a star with magnitude $H$ lies on two APOGEE fields and we have the selection functions $\operatorname{S}_1(H)$ and $\operatorname{S}_2(H)$ for those two fields, then the probability that the star was selected in APOGEE is the union of those two probabilities,
\begin{equation}
    \operatorname{S}(H) = \operatorname{S}_1(H) + \operatorname{S}_2(H) - \operatorname{S}_1(H)\operatorname{S}_2(H)
\end{equation}
if and only if the selection of the star on fields 1 and 2 are independent. In general, the selection probability for a star lying on $N$ independent fields is the union of the selection probabilities in each field. This formulation allows Model C to avoid `discovering' the overlaps between fields from the data. We note that the assumption that each field independently selects sources is likely not perfect, because the targeting within each field will have purposefully avoided targeting a star already scheduled for observations within another field.

We select all stars in APOGEE and 2MASS that satisfy $7 < H < 13.8\;\mathrm{mag}$ and that lie inside each of the APOGEE fields. We count the stars in each field in magnitude bins in APOGEE $k_{pm}$ and 2MASS $n_{pm}$, using $0.1\;\mathrm{mag}$ bins over $7<H<13.8\;\mathrm{mag}$, ensuring that we only include stars if they are flagged as being observed in that field. We assume that the mean of the model is $\mu=-10$ and that the magnitude dependence is described by the addition of two Rational Quadratic kernels, one with a variance of $\sigma^2=3$, a lengthscale of $0.2\;\mathrm{mag}$ and a powerlaw of $\alpha=1$ and the other with a variance of $\sigma^2=1$, a lengthscale of $1.0\;\mathrm{mag}$ and a powerlaw of $\alpha=1$.

\begin{figure}
    \centering
	\includegraphics[width=\linewidth]{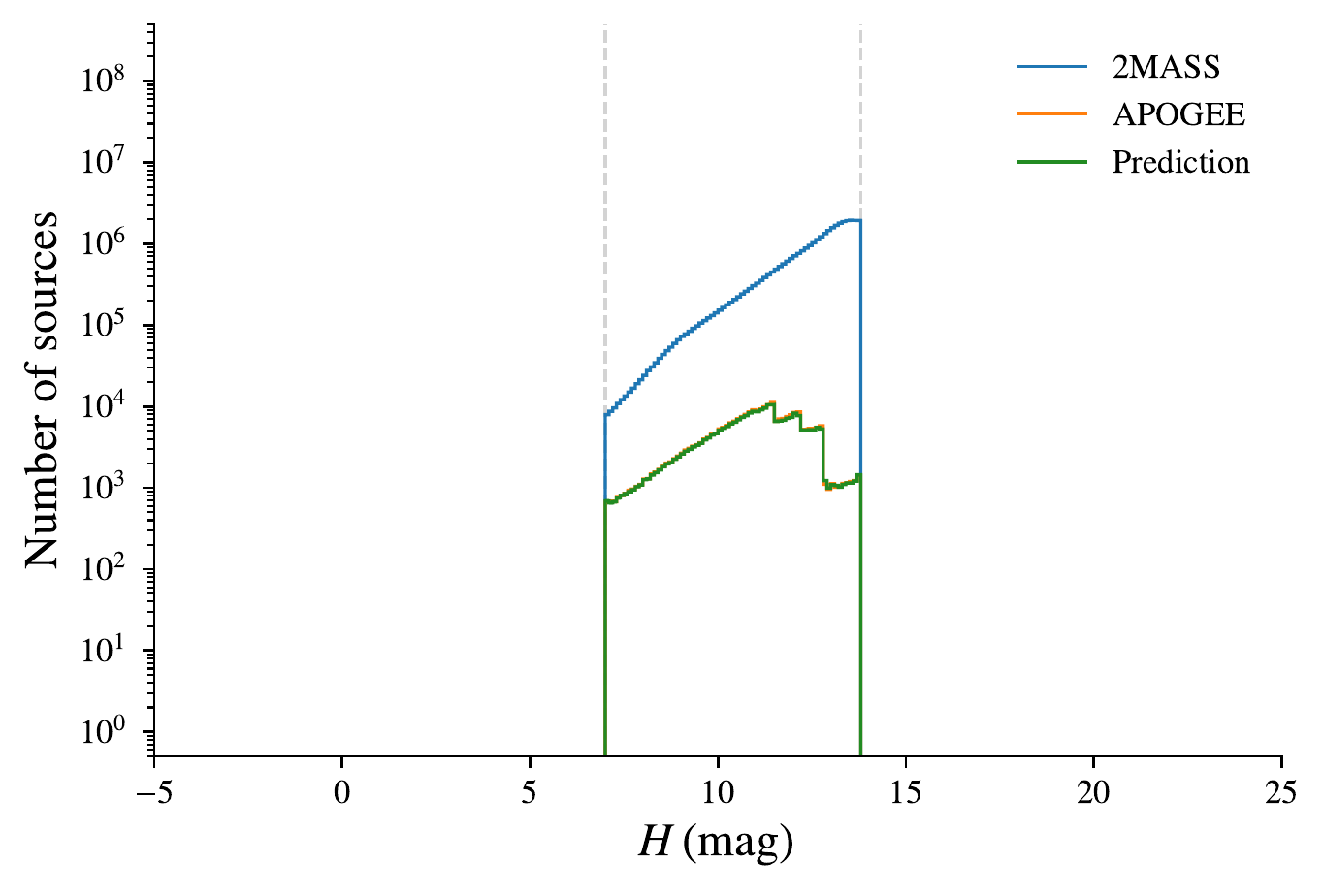}
	\caption{Same as Fig. \ref{fig:naive_magnitude_selection} but for Model C. The right grey line has been moved to $H=13.8\;\mathrm{mag}$, which is the faintest limit of the sample.}
	\label{fig:basic_magnitude_selection}
\end{figure}

We then computed the overall APOGEE selection probability at the centres of a $\textsc{nside}=1024$ HEALPix grid in steps of $0.1\;\mathrm{mag}$ across $7 < H < 13.8\;\mathrm{mag}$. The overall selection probability is the union of the selection probabilities in each field that contains the pixel centre. To evaluate the efficacy of our selection function (and to create plots comparable to those of Models A and B) we count the number of stars in APOGEE and 2MASS in each of these position-magnitude bins. We download the data at $\textsc{nside}=128$ resolution and average our inferred overall APOGEE selection probability maps down to that resolution. In Fig. \ref{fig:basic_magnitude_selection} we show the total number of 2MASS and APOGEE stars in each $0.1\;\mathrm{mag}$ bin, as well as a prediction for the number of APOGEE stars given the stars in 2MASS and our inferred selection function. We recover the magnitude distribution of the APOGEE sources with a fair degree of accuracy. The performance is worse than Model B (see Fig. \ref{fig:parent_magnitude_selection}) because Model C attempts to infer the magnitude-dependence of the selection probability independently in each field, while Model B is able to share knowledge of the magnitude-dependence between nearby fields.

\begin{figure*}
    \centering
	\includegraphics[width=\linewidth]{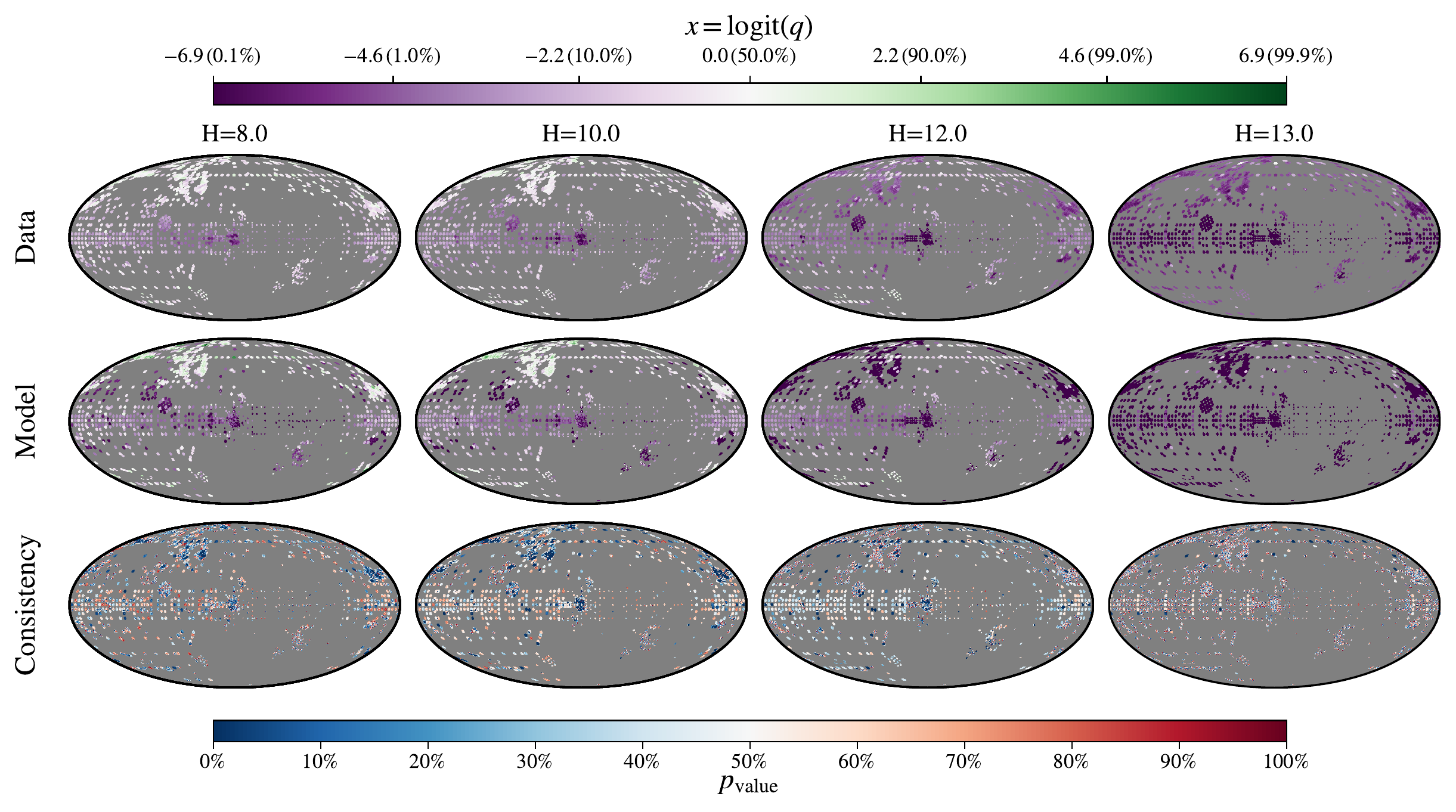}
	\caption{Same as Fig. \ref{fig:naive_position_selection} but for Model C. A mask has been applied to grey out regions outside APOGEE fields.}
	\label{fig:basic_position_selection}
\end{figure*}

In Fig. \ref{fig:basic_position_selection} we show the performance of our model across the sky. To arrive at a figure that is comparable to Figs. \ref{fig:naive_position_selection} and \ref{fig:parent_position_selection}, we have attributed to each pixel the total number of stars in APOGEE and 2MASS in every field containing that pixel. The motivation for this is that our simple $(1+k)/(2+n)$ estimate of the selection function should be applied across the entirety of each field, so that it is consistent with our assumption in Model C that the selection function is position-independent within each field. The position selection function is broadly consistent with the data across the entire sky.

\subsection{Summary}
Each of the three models presented above takes a different approach to modelling the APOGEE DR16 red giant selection function. Here we give an overview of the advantages and weaknesses of each of those approaches, before discussing a possible improvement that was beyond the scope of this paper to implement.

Model A approaches the problem from a position of complete ignorance about how the APOGEE collaboration carried out their targeting, ignoring that stars were only selected in specific fields on the sky and in a specific magnitude range. Ignoring that information inevitably leads to a worse estimate of the selection function, with the model having to smooth out the sharp drop in selection probability across these boundaries. However, this situation of complete ignorance is a common one when attempting to estimate the selection function of a subset of a catalogue, either because the selection of the subset was less well documented than the APOGEE targeting or because the selection cannot be simply expressed in terms of position, magnitude and colour. Model A thus gives a template for estimating the selection function of any subset selected from any catalogue, only requiring that we know the number of stars in the subset and catalogue in bins across position, magnitude and colour.

Model B assumes that 2MASS stars outside the APOGEE fields and the magnitude range $7\leq H \leq 13.8\;\mathrm{mag}$ cannot have been selected and so drops those stars from the star counts used in the modelling. It is incumbent on us to implement those position and magnitude cuts whenever we use the Model B selection function estimate, because \textsc{selectionfunctiontoolbox} will only be modelling the residual unknown component of the selection function. Incorporating knowledge of these cuts obviates the need for the model to smooth out the selection probability across these sharp discontinuities, resulting in a much better estimate of the selection function. Model B is the ideal model to use for a sample selected in specific fields that do not overlap and where it is possible that the selection probability could vary across a field. The latter of these assumptions does apply to APOGEE: the selection probability is lower on the side of the field nearer the Galactic plane due to there simply being more stars in the catalogue to pick from. However, many of the APOGEE fields do overlap with other fields, as shown in Fig. \ref{fig:apogee_overlap}. Stars in the 2MASS catalogue that lie in these overlaps are more likely to be in the APOGEE sample, because they had more opportunities to be targeted for observations. The overlapping fields create regions of the sky where the selection probability changes sharply which are too small for Model B to resolve.

Model C explores one solution to the overlapping fields problem, which is to separately model the probability of a 2MASS star being selected in each field as a function solely of magnitude. The probability of a star being selected in the overall APOGEE survey can then be computed as the union of the probabilities of it being selected in each of the fields containing that star, allowing us to perfectly resolve arbitrarily small overlaps between APOGEE fields.  The downside to this approach is that we must individually infer the selection probability in terms of magnitude in each field using only the stars in that field, ignoring that the magnitude selection in nearby fields will likely be similar and thus arrive at a worse estimate.

The ideal model would combine Model B and Model C. One solution would be to decompose the probability $q_{pmc}^{i}$ that a star is targeted by field $i$ into two parts: the indicator function returning one if the star lies in field $i$ and zero otherwise, and a probability $q_{pmc}$ that only depends on position, magnitude and colour and is common between fields. The probability of a star being selected in the overall APOGEE survey can then be calculated as in Model C, using the union of the selection probabilities in each field. This model correlates the probability of selection in terms of magnitude between fields and properly accounts for the overlap. Furthermore, it is easy to implement. If we have the binned count of stars in field $i$ in 2MASS $n_{pmc}^i$ and APOGEE $k_{pmc}^i$, then the quantities $n_{pmc} = \sum_i n_{pmc}^i$ and $k_{pmc} = \sum_i k_{pmc}^i$ can be immediately used with our standard spherical harmonic or spherical needlet machinery to derive an estimate of $q_{pmc}$. The difference between this model and Models A and B is that each star is counted multiple times, once for each APOGEE field in which it lies. This multiple-counting correctly accounts for the increased selection probability of a star sitting in the overlap of multiple fields. However, the assumption that the selection probability is solely a function of position, magnitude and colour that is common across fields is likely only partly true. This assumption would be violated by, for instance, two nearby fields having different magnitude limits. Examples of this can be seen in the right two panels of the top row of Fig. \ref{fig:naive_position_selection}, where only some of the fields in the northern hemisphere went down to $H=13\;\mathrm{mag}$. If fields with different magnitude limits are separated by more than a few degrees then the model will have enough flexibility to recover those magnitude limits. Otherwise, the model will average out the different selection probabilities, giving results that will still be somewhat accurate. In the worst case, suppose two overlapping fields have true selection probabilities of one and zero (i.e. one field selects all 2MASS sources in that field, the other selects none). The model will return a selection probability in each field of roughly 50\% in the overlap, and thus give an overall selection probability of 75\% once the union of probabilities is calculated, as compared to the true overall selection probability of 100\%.

\section{Conclusions}
\label{sec:cthen the flexonclusion}

Large catalogues are now ubiquitous throughout astronomy, but it remains true that most scientific analysis are done on smaller samples drawn from these catalogues by applying custom cuts. The selection function of that scientific sample - the probability that a star in the catalogue will satisfy these cuts and so make it into the sample - is thus unique to each scientific analysis. We have created a general framework that can flexibly estimate the selection function of a sample drawn from a catalogue in terms of position, magnitude and colour, accounting for correlations in the selection function using Gaussian processes and spherical harmonics. We have created a new open-source Python package \textsc{selectionfunctiontoolbox} that implements this framework and used it to estimate three different selection functions for the APOGEE DR16 red giant sample as a subset of 2MASS, with each selection function improving on the last by incorporating more domain knowledge of the APOGEE targeting. In a companion paper we applied our methodology to derive estimates of the astrometric and spectroscopic selection functions of \gaia EDR3. Our framework will make it trivial for astrophysicists to estimate the selection function that they should be using with the custom sample of stars that they designed to answer their scientific question.

\section*{Acknowledgements}
DB thanks Magdalen College for his fellowship and the Rudolf Peierls Centre for Theoretical Physics for providing office space and travel funds. AE thanks the Science and Technology Facilities Council of
the United Kingdom for financial support. The authors thank Payel Das for an in-depth peer review that helped us to greatly improve our original manuscript. This research has made use of the VizieR catalogue access tool, CDS,
 Strasbourg, France \citep{Ochsenbein2000}. This publication makes use of data products from the Two Micron All Sky Survey, which is a joint project of the University of Massachusetts and the Infrared Processing and Analysis Center/California Institute of Technology, funded by the National Aeronautics and Space Administration and the National Science Foundation. Funding for the Sloan Digital Sky 
Survey IV has been provided by the 
Alfred P. Sloan Foundation, the U.S. 
Department of Energy Office of 
Science, and the Participating 
Institutions. SDSS-IV acknowledges support and 
resources from the Center for High 
Performance Computing  at the 
University of Utah. The SDSS 
website is \url{www.sdss.org}. SDSS-IV is managed by the 
Astrophysical Research Consortium 
for the Participating Institutions 
of the SDSS Collaboration including 
the Brazilian Participation Group, 
the Carnegie Institution for Science, 
Carnegie Mellon University, Center for 
Astrophysics | Harvard \& 
Smithsonian, the Chilean Participation 
Group, the French Participation Group, 
Instituto de Astrof\'isica de 
Canarias, The Johns Hopkins 
University, Kavli Institute for the 
Physics and Mathematics of the 
Universe (IPMU) / University of 
Tokyo, the Korean Participation Group, 
Lawrence Berkeley National Laboratory, 
Leibniz Institut f\"ur Astrophysik 
Potsdam (AIP),  Max-Planck-Institut 
f\"ur Astronomie (MPIA Heidelberg), 
Max-Planck-Institut f\"ur 
Astrophysik (MPA Garching), 
Max-Planck-Institut f\"ur 
Extraterrestrische Physik (MPE), 
National Astronomical Observatories of 
China, New Mexico State University, 
New York University, University of 
Notre Dame, Observat\'ario 
Nacional / MCTI, The Ohio State 
University, Pennsylvania State 
University, Shanghai 
Astronomical Observatory, United 
Kingdom Participation Group, 
Universidad Nacional Aut\'onoma 
de M\'exico, University of Arizona, 
University of Colorado Boulder, 
University of Oxford, University of 
Portsmouth, University of Utah, 
University of Virginia, University 
of Washington, University of 
Wisconsin, Vanderbilt University, 
and Yale University.

\section*{Data availability}
The data underlying this article are publicly available on VizieR. All of the code underlying the figures is available in the Examples directory of the \textsc{selectionfunctiontoolbox} GitHub page: \url{https://github.com/gaiaverse/selectionfunctiontoolbox}.




\bibliographystyle{mnras}
\bibliography{references} 

\begin{thebibliography}{}
\makeatletter
\relax
\def\mn@urlcharsother{\let\do\@makeother \do\$\do\&\do\#\do\^\do\_\do\%\do\~}
\def\mn@doi{\begingroup\mn@urlcharsother \@ifnextchar [ {\mn@doi@}
  {\mn@doi@[]}}
\def\mn@doi@[#1]#2{\def\@tempa{#1}\ifx\@tempa\@empty \href
  {http://dx.doi.org/#2} {doi:#2}\else \href {http://dx.doi.org/#2} {#1}\fi
  \endgroup}
\def\mn@eprint#1#2{\mn@eprint@#1:#2::\@nil}
\def\mn@eprint@arXiv#1{\href {http://arxiv.org/abs/#1} {{\tt arXiv:#1}}}
\def\mn@eprint@dblp#1{\href {http://dblp.uni-trier.de/rec/bibtex/#1.xml}
  {dblp:#1}}
\def\mn@eprint@#1:#2:#3:#4\@nil{\def\@tempa {#1}\def\@tempb {#2}\def\@tempc
  {#3}\ifx \@tempc \@empty \let \@tempc \@tempb \let \@tempb \@tempa \fi \ifx
  \@tempb \@empty \def\@tempb {arXiv}\fi \@ifundefined
  {mn@eprint@\@tempb}{\@tempb:\@tempc}{\expandafter \expandafter \csname
  mn@eprint@\@tempb\endcsname \expandafter{\@tempc}}}

\bibitem[\protect\citeauthoryear{{Ahumada} \& et al.}{{Ahumada} \&
  et~al.}{2020}]{Ahumada2020}
{Ahumada} R.,  et al. 2020, \mn@doi [\apjs] {10.3847/1538-4365/ab929e}, \href
  {https://ui.adsabs.harvard.edu/abs/2020ApJS..249....3A} {249, 3}

\bibitem[\protect\citeauthoryear{{Blanton} \& et al.}{{Blanton} \&
  et~al.}{2017}]{Blanton2017}
{Blanton} M.~R.,  et al. 2017, \mn@doi [\aj] {10.3847/1538-3881/aa7567}, \href
  {https://ui.adsabs.harvard.edu/abs/2017AJ....154...28B} {154, 28}

\bibitem[\protect\citeauthoryear{{Boubert} \& {Everall}}{{Boubert} \&
  {Everall}}{2020}]{PaperII}
{Boubert} D.,  {Everall} A.,  2020, \mn@doi [\mnras] {10.1093/mnras/staa2305},
  \href {https://ui.adsabs.harvard.edu/abs/2020MNRAS.497.4246B} {497, 4246}

\bibitem[\protect\citeauthoryear{{Boubert}, {Everall}  \& {Holl}}{{Boubert}
  et~al.}{2020}]{PaperI}
{Boubert} D.,  {Everall} A.,   {Holl} B.,  2020, \mn@doi [\mnras]
  {10.1093/mnras/staa2050}, \href
  {https://ui.adsabs.harvard.edu/abs/2020MNRAS.497.1826B} {497, 1826}

\bibitem[\protect\citeauthoryear{{Boubert}, {Everall}, {Fraser}, {Gration}  \&
  {Holl}}{{Boubert} et~al.}{2021}]{PaperIII}
{Boubert} D.,  {Everall} A.,  {Fraser} J.,  {Gration} A.,   {Holl} B.,  2021,
  \mn@doi [\mnras] {10.1093/mnras/staa3791}, \href
  {https://ui.adsabs.harvard.edu/abs/2021MNRAS.501.2954B} {501, 2954}

\bibitem[\protect\citeauthoryear{{Bovy}, {Rix}, {Liu}, {Hogg}, {Beers}  \&
  {Lee}}{{Bovy} et~al.}{2012}]{Bovy2012Segue}
{Bovy} J.,  {Rix} H.-W.,  {Liu} C.,  {Hogg} D.~W.,  {Beers} T.~C.,   {Lee}
  Y.~S.,  2012, \mn@doi [\apj] {10.1088/0004-637X/753/2/148}, \href
  {https://ui.adsabs.harvard.edu/abs/2012ApJ...753..148B} {753, 148}

\bibitem[\protect\citeauthoryear{{Bovy} et~al.,}{{Bovy}
  et~al.}{2014}]{Bovy2014}
{Bovy} J.,  et~al., 2014, \mn@doi [\apj] {10.1088/0004-637X/790/2/127}, \href
  {https://ui.adsabs.harvard.edu/abs/2014ApJ...790..127B} {790, 127}

\bibitem[\protect\citeauthoryear{{Carpenter} et~al.,}{{Carpenter}
  et~al.}{2017}]{Carpenter2017}
{Carpenter} B.,  et~al., 2017, Journal of Statistical Software, \href
  {https://ui.adsabs.harvard.edu/abs/2017JSS....76....1C} {76, 1}

\bibitem[\protect\citeauthoryear{{Chen}, {Liu}, {Yuan}, {Xiang}, {Huang},
  {Wang}, {Zhang}  \& {Tian}}{{Chen} et~al.}{2018}]{Chen2018}
{Chen} B.~Q.,  {Liu} X.~W.,  {Yuan} H.~B.,  {Xiang} M.~S.,  {Huang} Y.,  {Wang}
  C.,  {Zhang} H.~W.,   {Tian} Z.~J.,  2018, \mn@doi [\mnras]
  {10.1093/mnras/sty454}, \href
  {https://ui.adsabs.harvard.edu/abs/2018MNRAS.476.3278C} {476, 3278}

\bibitem[\protect\citeauthoryear{{Das} \& {Binney}}{{Das} \&
  {Binney}}{2016}]{Das2016a}
{Das} P.,  {Binney} J.,  2016, \mn@doi [\mnras] {10.1093/mnras/stw744}, \href
  {https://ui.adsabs.harvard.edu/abs/2016MNRAS.460.1725D} {460, 1725}

\bibitem[\protect\citeauthoryear{{Das}, {Williams}  \& {Binney}}{{Das}
  et~al.}{2016}]{Das2016b}
{Das} P.,  {Williams} A.,   {Binney} J.,  2016, \mn@doi [\mnras]
  {10.1093/mnras/stw2167}, \href
  {https://ui.adsabs.harvard.edu/abs/2016MNRAS.463.3169D} {463, 3169}

\bibitem[\protect\citeauthoryear{{Everall} \& {Das}}{{Everall} \&
  {Das}}{2020}]{Everall2020}
{Everall} A.,  {Das} P.,  2020, \mn@doi [\mnras] {10.1093/mnras/staa283}, \href
  {https://ui.adsabs.harvard.edu/abs/2020MNRAS.493.2042E} {493, 2042}

\bibitem[\protect\citeauthoryear{{Everall}, {Boubert}, {Koposov}, {Smith}  \&
  {Holl}}{{Everall} et~al.}{2021}]{PaperIV}
{Everall} A.,  {Boubert} D.,  {Koposov} S.~E.,  {Smith} L.,   {Holl} B.,  2021,
  \mn@doi [\mnras] {10.1093/mnras/stab041}, \href
  {https://ui.adsabs.harvard.edu/abs/2021MNRAS.502.1908E} {502, 1908}

\bibitem[\protect\citeauthoryear{Geller \& Mayeli}{Geller \&
  Mayeli}{2010}]{Geller2009}
Geller D.,  Mayeli A.,  2010, Theory Signal Image Process, 9, 1

\bibitem[\protect\citeauthoryear{{G{\'o}rski}, {Hivon}, {Banday}, {Wand elt},
  {Hansen}, {Reinecke}  \& {Bartelmann}}{{G{\'o}rski}
  et~al.}{2005}]{Gorski2005}
{G{\'o}rski} K.~M.,  {Hivon} E.,  {Banday} A.~J.,  {Wand elt} B.~D.,  {Hansen}
  F.~K.,  {Reinecke} M.,   {Bartelmann} M.,  2005, \mn@doi [\apj]
  {10.1086/427976}, \href
  {https://ui.adsabs.harvard.edu/abs/2005ApJ...622..759G} {622, 759}

\bibitem[\protect\citeauthoryear{Harbrecht, Peters  \& Schneider}{Harbrecht
  et~al.}{2012}]{Harbrecht2012}
Harbrecht H.,  Peters M.,   Schneider R.,  2012, Applied numerical mathematics,
  62, 428

\bibitem[\protect\citeauthoryear{Liu \& Nocedal}{Liu \&
  Nocedal}{1989}]{Liu1989}
Liu D.~C.,  Nocedal J.,  1989, Mathematical programming, 45, 503

\bibitem[\protect\citeauthoryear{{Mackereth} \& {Bovy}}{{Mackereth} \&
  {Bovy}}{2020}]{Mackereth2020}
{Mackereth} J.~T.,  {Bovy} J.,  2020, \mn@doi [\mnras] {10.1093/mnras/staa047},
  \href {https://ui.adsabs.harvard.edu/abs/2020MNRAS.492.3631M} {492, 3631}

\bibitem[\protect\citeauthoryear{{Majewski} et~al.,}{{Majewski}
  et~al.}{2017}]{Majewski2017}
{Majewski} S.~R.,  et~al., 2017, \mn@doi [\aj] {10.3847/1538-3881/aa784d},
  \href {https://ui.adsabs.harvard.edu/abs/2017AJ....154...94M} {154, 94}

\bibitem[\protect\citeauthoryear{{Mints} \& {Hekker}}{{Mints} \&
  {Hekker}}{2019}]{Mints2019}
{Mints} A.,  {Hekker} S.,  2019, \mn@doi [\aap] {10.1051/0004-6361/201834256},
  \href {https://ui.adsabs.harvard.edu/abs/2019A&A...621A..17M} {621, A17}

\bibitem[\protect\citeauthoryear{{Nandakumar}, {Schultheis}, {Hayden},
  {Rojas-Arriagada}, {Kordopatis}  \& {Haywood}}{{Nandakumar}
  et~al.}{2017}]{Nandakumar2017}
{Nandakumar} G.,  {Schultheis} M.,  {Hayden} M.,  {Rojas-Arriagada} A.,
  {Kordopatis} G.,   {Haywood} M.,  2017, \mn@doi [\aap]
  {10.1051/0004-6361/201731099}, \href
  {https://ui.adsabs.harvard.edu/abs/2017A&A...606A..97N} {606, A97}

\bibitem[\protect\citeauthoryear{{Nidever} et~al.,}{{Nidever}
  et~al.}{2015}]{Nidever2015}
{Nidever} D.~L.,  et~al., 2015, \mn@doi [\aj] {10.1088/0004-6256/150/6/173},
  \href {https://ui.adsabs.harvard.edu/abs/2015AJ....150..173N} {150, 173}

\bibitem[\protect\citeauthoryear{{Ochsenbein}, {Bauer}  \&
  {Marcout}}{{Ochsenbein} et~al.}{2000}]{Ochsenbein2000}
{Ochsenbein} F.,  {Bauer} P.,   {Marcout} J.,  2000, \mn@doi [\aaps]
  {10.1051/aas:2000169}, \href
  {https://ui.adsabs.harvard.edu/abs/2000A&AS..143...23O} {143, 23}

\bibitem[\protect\citeauthoryear{Rasmussen \& Williams}{Rasmussen \&
  Williams}{2006}]{rasmussen2006}
Rasmussen C.~E.,  Williams C. K.~I.,  2006, Gaussian processes for machine
  learning.
MIT Press, Cambridge

\bibitem[\protect\citeauthoryear{{Reinecke}}{{Reinecke}}{2011}]{Reinecke2011}
{Reinecke} M.,  2011, \mn@doi [\aap] {10.1051/0004-6361/201015906}, \href
  {https://ui.adsabs.harvard.edu/abs/2011A&A...526A.108R} {526, A108}

\bibitem[\protect\citeauthoryear{{Rix} et~al.,}{{Rix} et~al.}{2021}]{Rix2021}
{Rix} H.-W.,  et~al., 2021, arXiv e-prints, \href
  {https://ui.adsabs.harvard.edu/abs/2021arXiv210607653R} {p. arXiv:2106.07653}

\bibitem[\protect\citeauthoryear{{Rybizki}, {Rix}, {Demleitner}, {Bailer-Jones}
   \& {Cooper}}{{Rybizki} et~al.}{2021}]{Rybizki2021}
{Rybizki} J.,  {Rix} H.-W.,  {Demleitner} M.,  {Bailer-Jones} C. A.~L.,
  {Cooper} W.~J.,  2021, \mn@doi [\mnras] {10.1093/mnras/staa3089}, \href
  {https://ui.adsabs.harvard.edu/abs/2021MNRAS.500..397R} {500, 397}

\bibitem[\protect\citeauthoryear{{Scodeller}, {Rudjord}, {Hansen}, {Marinucci},
  {Geller}  \& {Mayeli}}{{Scodeller} et~al.}{2011}]{Scodeller2011}
{Scodeller} S.,  {Rudjord} {\O}.,  {Hansen} F.~K.,  {Marinucci} D.,  {Geller}
  D.,   {Mayeli} A.,  2011, \mn@doi [\apj] {10.1088/0004-637X/733/2/121}, \href
  {https://ui.adsabs.harvard.edu/abs/2011ApJ...733..121S} {733, 121}

\bibitem[\protect\citeauthoryear{{Skrutskie} et~al.,}{{Skrutskie}
  et~al.}{2006}]{Skrutskie2006}
{Skrutskie} M.~F.,  et~al., 2006, \mn@doi [\aj] {10.1086/498708}, \href
  {https://ui.adsabs.harvard.edu/abs/2006AJ....131.1163S} {131, 1163}

\bibitem[\protect\citeauthoryear{{Stonkut{\.{e}}} et~al.,}{{Stonkut{\.{e}}}
  et~al.}{2016}]{Stonkute2016}
{Stonkut{\.{e}}} E.,  et~al., 2016, \mn@doi [\mnras] {10.1093/mnras/stw1011},
  \href {https://ui.adsabs.harvard.edu/abs/2016MNRAS.460.1131S} {460, 1131}

\bibitem[\protect\citeauthoryear{{Wilson} et~al.,}{{Wilson}
  et~al.}{2019}]{Wilson2019}
{Wilson} J.~C.,  et~al., 2019, \mn@doi [\pasp] {10.1088/1538-3873/ab0075},
  \href {https://ui.adsabs.harvard.edu/abs/2019PASP..131e5001W} {131, 055001}

\bibitem[\protect\citeauthoryear{{Wojno} et~al.,}{{Wojno}
  et~al.}{2017}]{Wojno2017}
{Wojno} J.,  et~al., 2017, \mn@doi [\mnras] {10.1093/mnras/stx606}, \href
  {https://ui.adsabs.harvard.edu/abs/2017MNRAS.468.3368W} {468, 3368}

\makeatother
\end{thebibliography}



\FloatBarrier

\appendix

\section{Avoiding overflow in the binomial log-likelihood}
\label{sec:overflow}

The logit-probability $x=\log\frac{p}{1-p}$ can take arbitrarily large values as $p$ goes towards zero or unity. To avoid numeric overflow when computing Eq. \ref{eq:loglikelihood}, we rearrange the naive expression,
\begin{align}
    \log\left(\cosh{x}\right) &= \log\left(\frac{e^{x}+e^{-x}}{2}\right) \\
    &= |x| + \log\left(\frac{1+e^{-2|x|}}{2}\right).
\end{align}
We note that the derivative of this expression can be similarly rearranged to save on computation by reusing the term $1+e^{-2|x|}$,
\begin{align}
    \tanh{x} &= \frac{e^{x}-e^{-x}}{e^{x}+e^{-x}} \\
    &= \operatorname{sign}(x)\frac{2-(1+e^{-2|x|})}{1+e^{-2|x|}}.
\end{align}

\section{Angular power spectrum}
\label{sec:powerspectrum}

Suppose we have a zero-mean, isotropic Gaussian random field $\Theta(\mathbf{x})$ at every point $\mathbf{x}$ on the sky. If we sample many realisations of that field and expand them in terms of spherical harmonics,
\begin{equation}
    \Theta(\mathbf{x}) = \sum_{\ell=0}^{\infty}\sum_{m=-\ell}^{\ell}a_{\ell m}Y_\ell^m(\mathbf{x})
\end{equation}
where
\begin{equation}
    a_{\ell m} = \int_{S^2}\Theta(\mathbf{x})\overline{Y}_\ell^m(\mathbf{x})\mathrm{d}\Omega,
\end{equation}
then we can use the variance of the resulting harmonic coefficients to define the power spectrum $C_\ell$ of the field by
\begin{equation}
    \langle a_{\ell m}\overline{a}_{\ell' m'}\rangle = \delta_{\ell\ell'}\delta_{mm'}C_\ell.
\end{equation}
In this work we are trying to forward model an unknown random field on the sphere in terms of random variables $a_{\ell m}$. If we know a reasonable a priori choice for the power spectrum of that field, then we can set the prior on each $a_{\ell m}$ to be a zero-mean Gaussian with a variance of $C_\ell$.

We can derive a similar expression for the case of needlets, where we are instead expanding the random field $\Theta(\mathbf{x})$ as
\begin{equation}
    \Theta(\mathbf{x}) = \sum_{j=0}^{\infty}\sum_{k=0}^{N_j}\beta_{jk}\psi_{jk}(\mathbf{x})
\end{equation}
where
\begin{align}
    \beta_{jk} &= \int_{S^2}\Theta(\mathbf{x})\psi_{jk}(\mathbf{x})\mathrm{d}\Omega \\
    &= \sqrt{\lambda_{jk}}\sum_{\ell=0}^{\infty}\operatorname{b}_\ell(j) \sum_{m=-\ell}^{\ell}Y_\ell^m(\xi_{jk})\int_{S^2}\Theta(\mathbf{x})\overline{Y}_\ell^m(\mathbf{x})\mathrm{d}\Omega \\
    &= \sqrt{\lambda_{jk}}\sum_{\ell=0}^{\infty}\operatorname{b}_\ell(j) \sum_{m=-\ell}^{\ell}a_{\ell m}Y_\ell^m(\xi_{jk}),
\end{align}
using the definition of the spherical needlets. The variance of the needlet coefficients over many realisations will be given by
\begin{align*}
    \langle &\beta_{jk}\overline{\beta}_{j'k'}\rangle \nonumber \\
    &=\left\langle \left(\sqrt{\lambda_{jk}}\sum_{\ell=0}^{\infty}\operatorname{b}_\ell(j) \sum_{m=-\ell}^{\ell}a_{\ell m}Y_\ell^m(\xi_{jk})\right)\right. \nonumber \\
    &\;\;\times\left.\left(\sqrt{\lambda_{j'k'}}\sum_{\ell'=0}^{\infty}\operatorname{b}_{\ell'}(j') \sum_{m'=-\ell'}^{\ell'}\overline{a}_{\ell' m'}\overline{Y}_{\ell'}^{m'}(\xi_{j'k'})\right) \right\rangle \\
    &= \sqrt{\lambda_{jk}}\sqrt{\lambda_{j'k'}}\sum_{\ell=0}^{\infty}\operatorname{b}_\ell(j)\operatorname{b}_\ell(j')C_\ell \sum_{m=-\ell}^{\ell} Y_\ell^m(\xi_{jk})\overline{Y}_{\ell}^{m}(\xi_{j'k'}) \\
    &= \sqrt{\lambda_{jk}}\sqrt{\lambda_{j'k'}}\sum_{\ell=0}^{\infty}\operatorname{b}_\ell(j)\operatorname{b}_\ell(j')C_\ell \frac{2\ell+1}{4\upi} \operatorname{P}_{\ell}(\langle\xi_{jk},\xi_{j'k'}\rangle), \\
\end{align*}
where $\langle\xi_{jk},\xi_{j'k'}\rangle$ is the inner product of these two unit vectors. This expression reveals that the random variables $\beta_{jk}$ are correlated between needlets of the same order and across orders. In our case we are simply looking for a prior for these random variables, and so we can focus on the case $j=j',k=k'$ to find a reasonable choice for the variance
\begin{equation}
    \langle |\beta_{jk}|^2\rangle = \frac{1}{N_j}\sum_{\ell=0}^{\infty} (2\ell+1)C_\ell \operatorname{b}_\ell^2(j).
\end{equation}

\section{Gaussian Process kernels}

\begin{table*}
	\centering
	\caption{The kernels that we have made available within \textsc{selectionfunctiontoolbox}.}
	\label{tab:kernels}
	\begin{tabular}{lcc} 
		\hline
		Name & Hyperparameters & Form\\
		\hline
		Squared Exponential & $l>0$ & $K(x,x'|l) = \exp{\left(-\frac{(x-x')^2}{2l^2}\right)}$\\
		Mat\'ern-1/2 & $l>0$ & $K(x,x'|l) = \exp{\left(-\frac{|x-x'|}{l}\right)}$\\
		Mat\'ern-3/2 & $l>0$ & $K(x,x'|l) = \left(1+\frac{\sqrt{3}|x-x'|}{l}\right)\exp{\left(-\frac{\sqrt{3}|x-x'|}{l}\right)}$\\
		Rational Quadratic & $l>0,\alpha>0$ & $K(x,x'|l,\alpha) = \left(1+\frac{(x-x')^2}{2\alpha l^2}\right)^{-\alpha}$\\
		Periodic & $l>0,p>0$ & $K(x,x'|l,p) = \exp{\left(-\frac{2\sin^2{(\pi|x-x'|/p)}}{l^2}\right)}$\\
		Linear & $-\infty<c<\infty$ & $K(x,x'|c) = (x-c)(x-c')$\\
		White & None & $K(x,x') = \begin{cases}1, \mathrm{if} x = x'\\ 0, \mathrm{otherwise} \end{cases}$\\
		\hline
		Scaled & $\sigma^2 \geq 0$, kernel $K_1$ & $K_{K_1}(x,x'|\sigma^2) = \sigma^2K_1(x,x')$\\
		Additive & kernels $K_1, K_2$ & $K_{K_1,K_2}(x,x') =  K_1(x,x') + K_2(x,x')$\\
		Multiplicative & kernels $K_1, K_2$ & $K_{K_1,K_2}(x,x') = K_1(x,x')K_2(x,x')$\\
		\hline
	\end{tabular}
\end{table*}

We have implemented a number of kernels in \textsc{selectionfunctiontoolbox} that can be used to describe the magnitude or colour dependence of selection functions, listed in the upper section of Tab. \ref{tab:kernels}. The act of scaling a kernel by a constant number, adding two kernels together or multipling two kernels all result in a new valid kernel, and so we have also implemented functionality allowing for these operations as listed in the lower section of Tab. \ref{tab:kernels}. The user is also free to implement their own kernels using our \textsc{kernelbase} class, allowing virtually any correlation between magnitude or colour bins to be encoded. The only requirement is that the resulting covariance matrix must be able to be decomposed into Cholesky factors, which requires that, at a minimum, the kernel must be real-valued, symmetric $K(x,x')=K(x',x)$, and satisfy $K(x,x)\neq 0$.


\bsp	
\label{lastpage}
\end{document}